\documentclass[review,5p]{elsarticle}
\journal{Journal of Crystal Growth}
\usepackage{hyperref}
\usepackage[utf8]{inputenc}
\usepackage{amsmath}
\usepackage{amssymb}
\DeclareMathOperator\erfc{erfc}
\usepackage{graphicx}
\usepackage[american]{babel}
\usepackage[amssymb]{SIunits}
\usepackage{subfigure}
\usepackage{notoccite}
\usepackage{afterpage}
\newcommand{\vmu}{\mbox{\boldmath{$\mu$}}}
\usepackage{tikz}
\usepackage{ulem}
\usetikzlibrary{calc}
\usetikzlibrary{plotmarks}
\usepackage{pgfplots}
\usepackage{etoolbox}
\patchcmd{\normalsize}{13.6}{13}{}{}
\pgfplotsset{
tick label style={font=\tiny},
label style={font=\small},
legend style={font=\tiny}
}

\begin{document}

\begin{frontmatter}
\title{Theoretical and numerical investigation of diffusive instabilities in multi-component alloys}

\author[add]{Arka Lahiri\corref{correspondingauthor}}

\ead{arka@platinum.materials.iisc.ernet.in}
 
\author[add]{Abhik Choudhury}%
\ead{abhiknc@materials.iisc.ernet.in}

\address[add]{%
Department of Materials Engineering, 
Indian Institute of Science, Bangalore - 560 012, India.
}%
\cortext[correspondingauthor]{Corresponding author}


\begin{abstract}
Mechanical properties of engineering alloys are strongly correlated to their 
microstructural length scale. Diffusive instabilities of the 
Mullins-Sekerka type is one of the principal mechanisms through which 
the scale of the microstructural features are determined during solidification. 
In contrast to binary systems, in multicomponent alloys with arbitrary 
interdiffusivities, the growth rate as well as the maximally growing 
wavelengths characterizing these instabilities depend on the the 
dynamically selected equilibrium tie-lines 
and the steady state growth velocity. In this study, we derive 
analytical expressions to characterize the dispersion behavior in 
isothermally solidified multicomponent (quaternary) alloys
for different choices of the inter-diffusivity matrices and confirm our calculations
using phase-field simulations. 
Thereafter, we perform controlled studies to capture and isolate the dependence of 
instability length scales on solute diffusivities and steady state planar 
front velocities, which leads to an understanding of the process of length 
scale selection during the onset of instability for any alloy 
composition with arbitrary diffusivities, comprising of
both independent and coupled diffusion of solutes. 
\end{abstract}

\end{frontmatter}


\section{Introduction} 
Morphological instability of a solid-liquid interface to small perturbations 
is the basis for the most 
commonly observed solidification microstructure of dendrites. 
Experimentally, a planar solidification front during solidification 
is usually perturbed, either by random thermal 
fluctuations or due to interactions with insoluble impurities~\cite{Kurz1986}. 
An unstable solidification front is characterized by 
amplification of such interfacial perturbations which ultimately develop into cellular or 
dendritic structures. Any random infinitesimal perturbation can be thought of as a linear combination 
of a multitude of wavelengths with different amplitudes (which are small during early stages). 
Thus, the stability of a solid-liquid interface to the amplification 
of these perturbations can be understood by investigating the growth behavior of the individual modes. 
Mullins and Sekerka in their classical work~\cite{Mullins1964} present 
a linear stability analysis of an interface perturbed by any generic wavelength 
and provide expressions for their growth rates for a binary alloy. 
This allows the determination of the maximally (fastest) growing wavelength, that principally 
can be related to the length scales in the cellular or dendritic microstructures. Cells and dendrites, 
 being the most commonly observed solidification structures, have been investigated
theoretically~\cite{Ivantsov1947,Horvay1961,Langer1978,Langer1980,Lipton1984,KurzP1986,Lipton1987,LiptonK1987}, 
experimentally~\cite{Weinberg1951,Weinberg1952,Lindmeyer1966,Glicksman1976,Somboonsuk1985,Arnold1999,Akamatsu1995,Akamatsu2016}, 
and as well as through simulations~\cite{Lan2004,Ramirez2004,Ramirez2005,Bottger2009,Plapp2010,Ma2014}. 


The theory in~\cite{Mullins1964} (also confirmed by phase-field simulations in~\cite{Echebarria2004}), reveals that the  
instability length scale in a binary alloy is a function of the equilibrium
compositions in the solid and the liquid, the composition of the supersaturated 
liquid (or equivalently the growth velocity in 
directionally solidified systems) and the solute diffusivities. 

For multicomponent 
alloys, the equilibrium compositions of the phases (or the effective tie-line) 
is a function of the inter-diffusivity matrix.
This selection of equilibrium tie-lines during growth introduces an additional degree of freedom 
which influences the behavior of the perturbations and thereby the length scales. 
This phenomenon of selection of tie-lines during growth of a planar interface  
has not been addressed by the theoretical discussions of morphological instability 
in directionally solidified multicomponent systems till now.
Among the earliest in this regard is the study performed by 
Coates et al.~\cite{Coates1968}, which is carried out in the context of a
ternary alloy with no diffusional interaction amongst solutes, with the 
dispersion behavior (the amplification rates for different 
wavelengths of perturbations) calculated assuming a steady state behavior 
in the perturbed state. The correctness of this assumption is investigated 
by Coriell et al.~\cite{Coriell1987} by solving for the time 
dependent problem which lead to the validation of the steady-state assumption 
in~\cite{Coates1968}. The effect of coupled solute diffusivities
on the stability of the system to infinitesimal perturbations is studied 
by Hunziker~\cite{Hunziker2001}, but without accounting for the 
possibility of a shift in the tie-lines playing a role in 
the selection of instability length scales.  

The fact that the diffusivity matrix in a multicomponent system 
influences the dynamic selection of tie-lines and growth velocities, 
has a considerable impact on the instability behavior.
Furthermore, in the context of the difficulties associated with experimental 
determination of diffusivities in multicomponent alloy systems, a theoretical or 
a numerical understanding of the instability behavior as a function of solute diffusivities
becomes even more important.
This motivates our study where we isolate the effect of each of these factors: 
diffusivity, steady-state (solidification by advancement of a planar front) growth velocity 
and tie-line compositions, to explain the problem of microstructural 
length scale selection in multicomponent systems displaying either independent or coupled diffusion of solutes.
In this paper, we derive an analytical theory and perform phase field simulations 
to establish all our major conclusions,
with the context being that of an isothermally solidifying system 
in contrast to the directional solidification studies mentioned in 
the preceding paragraphs. We start with the theoretical analysis of the 
growth of perturbations and thereafter 
describe the phase-field model, followed by the results.

\section{Theory}
We begin with steady-state (planar front solidification) of a $K$ component alloy.
The $K-1$ independent components have no diffusional interaction 
(i.e., the diffusivity matrix is diagonal) in the liquid, while there is no diffusion in the solid. 
The governing equation in a frame attached to the interface growing at a velocity $V$~\footnote{the value which is 
asymptotically approached, late into the scaling regime.} is,

\begin{align}
 D_{ii}\frac{\partial^2 c_i}{\partial z^2} + V \frac{\partial c_i}{\partial z} = 0,
 \label{planar_gde}
\end{align}
where $c_i$ denotes the concentration and $D_{ii}$ the diffusivity of the 
$i$'th component in the liquid, with $i = 1, 2, 3, \ldots, K-1$. $z$ is the direction 
normal to the solid-liquid interface (located at $z=0$). Consideration of uncoupled 
diffusion of solutes, enables us to present the following discussion in terms of 
a generic component $i$, which stands for all the components in a system. 

The solution to Eq.~\ref{planar_gde} has to obey the following boundary conditions:
\begin{align}
 c_i = c_{i,eq}^l,\  \textrm{at} \  z=0,
 \label{planar_bc_1}
\end{align}
which is the equilibrium composition in the liquid given
by the tie-line selected during growth, and, 
\begin{align}
 V c_{i,eq}^l (1-k_i)=-D_{ii}\frac{\partial c_i}{\partial z} \Bigg|_{z=0}=-D_{ii} G_i,
 \label{planar_bc_2}
\end{align}
which is the Stefan boundary condition at a solid-liquid 
interface moving with velocity $V$. $G_i$ is the
gradient in $c_i$ at the planar interface. $k_i$ is the equilibrium
partition coefficient corresponding to the selected tie-line. 
The solution to Eq.~\ref{planar_gde}
which conforms to the boundary conditions in Eqs.~\ref{planar_bc_1} and~\ref{planar_bc_2}, is given by,
\begin{align}
c_i=c_{i,eq}^l + \frac{G_i D_{ii}}{V}\left[ 1 - \exp \left(\frac{-Vz}{D_{ii}} \right) \right].
\label{planar_sol}
\end{align}

The steady-state solidification described above is now modified by introducing a 
sinusoidal perturbation given by, 
\begin{align}
 z=\Phi=\delta (t) \sin \omega x,
 \label{pert_eq}
\end{align}
with $x$ being one of the directions parallel to the unperturbed interface (normal to $z$). 
Despite $\delta$ being a function of time ($t$), a stability
criterion derivable from the steady state solution will not differ appreciably 
from that obtained by solving the time dependent problem~\cite{Coates1968,Coriell1987}, 
which leads to the following governing differential equation describing a system with interfacial perturbations,
\begin{align}
D_{ii}\frac{\partial^2 \widetilde{c_i}}{\partial z^2}+ D_{ii}\frac{\partial^2 \widetilde{c_i}}{\partial x^2} + 
V \frac{\partial \widetilde{c_i}}{\partial z} =0,
\label{pert_gde}
\end{align}
where the modified composition field of any generic component $i$, under 
interfacial perturbation is denoted by
$\widetilde{c_i}$. The form of the solution to Eq.~\ref{pert_gde}, is obtained 
by adding a term to the steady-state solution given by Eq.~\ref{planar_sol}, which 
represents a sinusoidal variation in the composition fields 
in response to the interfacial perturbation of a similar character.
It must be taken into account that such an effect diminishes in magnitude with 
distance from the interface, leading to the following expression,

\begin{align}
 \widetilde{c_i} &= c_i+E_i \sin \omega x \exp \left(-k_\omega^{(i)} z \right) \nonumber\\ 
 &= c_{i,eq}^l + \frac{G_i D_{ii}}{V}\left[ 1 - \exp \left(\frac{-Vz}{D_{ii}} \right) \right]\nonumber \\
 &+E_i \sin \omega x \exp \left(-k_\omega^{(i)} z \right),
 \label{pert_sol_form}
\end{align}

where $k_\omega^{(i)}$ and $E_i$ are constants. 
The constant $k_\omega^{(i)}$ is determined by the requirement
that the composition profile given by Eq.~\ref{pert_sol_form} 
satisfies the governing Eq.~\ref{pert_gde}, resulting in a
quadratic equation in $k_\omega^{(i)}$, which yields,
\begin{align}	
  k_\omega^{(i)}=\frac{V}{2D_{ii}} + \sqrt{{\left(\frac{V}{2D_{ii}}\right)}^2 + \omega^2}.
  \label{komega}
\end{align}
  
The compositions in the liquid at the perturbed interface are no longer given by the 
equilibrium tie-lines selected during steady-state growth because of the Gibbs-Thomson
correction. The composition deviations conform to the interfacial curvature, which 
is approximated by the second derivative of $z$ with respect to $x$ from Eq.~\ref{pert_eq}
and can be seen to be of the same form as the perturbation itself. Thus, the composition in
the liquid at the perturbed interface is given by,
\begin{align}
c_{i,\Phi}=c_{i,eq}^l + b_{i} \delta \sin \omega x,
\label{pert_interf}
\end{align}
where $b_i$ is a constant. 
Evaluating the solution to the perturbed problem given by Eq.~\ref{pert_sol_form} 
at the perturbed interface (see Eq.~\ref{pert_eq}), we retrieve,
\begin{align}
c_{i, \Phi} = c_{i,eq}^l + \left(G_i \delta + E_i \right) \sin \omega x - \delta  k_\omega^{(i)} E_i \sin^2 \omega x.
\label{pert_interf_2}
\end{align}
 Separately comparing the Fourier coefficients and the leading order constant
 from Eq.~\ref{pert_interf} and Eq.~\ref{pert_interf_2}, we derive,
\begin{align}
 E_i=\delta (b_i-G_i).
 \label{const_rel}
\end{align}
Eq.~\ref{const_rel} is only a reformulation of $E_i$ in terms of $b_i$, which is still unknown. 
The $b_i$'s ($b_1$, $b_2$, $\cdots$, $b_{k-1}$) are related to each other through the fact that
each of the composition fields ($\widetilde{c_i}$) satisfies the
Stefan condition at the perturbed interface, moving at a velocity ($v(x)$).
This implies that the same amplification factor ($\dot{\delta}/\delta$) 
must be obtained by considering the diffusion field of any one of the components. 
The expression for the Stefan condition at the perturbed interface, is given by,
\begin{align}
v(x) &= \left( V + \dot{\delta} \sin \omega x \right) c_{i, \Phi} \left(1 - k_i \right) \nonumber \\ 
 &= -D_{ii} \frac{\partial \widetilde{c_i}}{\partial z} \Bigg{|}_{z=\delta \sin \omega x},
 \label{pert_stefan}
\end{align}
where $\dot{\delta}$ is $d\delta/dt$. The above equation can be re-expressed as:
\begin{align}
 \left( V + \dot{\delta} \sin \omega x \right)=\frac{-D_{ii}}{c_{i, \Phi}
 \left(1 - k_i \right)}  \frac{\partial \widetilde{c_i}}{\partial z} \Bigg{|}_{z=\delta \sin \omega x}.
 \label{pert_stefan_2}
\end{align}

From Eq.~\ref{pert_interf},
\begin{align}
 \frac{1}{c_{i, \Phi}} &= \frac{1}{c_{i,eq}^l\left(1+\frac{b_i}{c_{i,eq}^l} \delta \sin \omega x\right)} \nonumber \\  
 &\approx \frac{1}{c_{i, eq}^l}\left(1-\frac{b_i}{c_{i, eq}^l} \delta \sin \omega x\right),
 \label{pert_int_comp_inv}
\end{align}
where, we have limited ourselves to terms linear in $\delta$. 
Employing Eq.~\ref{const_rel} in the expression obtained  by differentiating Eq.~\ref{pert_sol_form} with respect to $z$,
we derive,
\begin{align}
\frac{\partial \widetilde{c_i}}{\partial z} \Bigg{|}_{z=\delta \sin \omega x} &= 
G_i\left(1-\frac{V\delta \sin \omega x}{D_{ii}}\right) \nonumber \\ &- k_\omega^{(i)} E_i  \sin \omega x (1-k_\omega^{(i)} \delta \sin \omega x) \nonumber \\
 &\approx G_i - \left( \frac{G_i V}{D_{ii}} + k_\omega^{(i)} (b_i-G_i) \right)\delta \sin \omega x,
\label{pert_int_grad}
\end{align}
by limiting ourselves to terms linear in $\delta$. Equating the Fourier coefficients from both sides of Eq.~\ref{pert_stefan_2}, we get:
\begin{align}
 \frac{\dot{\delta}}{\delta}=
 V\widetilde{\omega_i}\left[-\frac{b_i}{G_i}+\frac{1}{\widetilde{\omega_i}}\left(k_\omega^{(i)}-\frac{V}{D_{ii}}\right)\right],
 \label{ampl_fac}
\end{align}
where,
\begin{align}
 \widetilde{\omega_i}=k_\omega^{(i)} - \frac{V}{D_{ii}}\left(1-k_i\right).
 \label{til_om}
\end{align}
Invoking the fact that $\dot{\delta}/\delta$ is a quantity unique to the system as a whole, regardless 
of the choice of component (i.e., $i$) in Eq.~\ref{ampl_fac}, leads to  
$K-2$ relations inter-relating the $b_i$'s ($b_1$, $b_2$, $\cdots$ $b_{k-1}$). 
To express all $b_i$'s ($i\neq 1$) in terms of $b_1$, we equate 
the algebraic expressions for $\dot{\delta}/\delta$ corresponding to each component, which writes as:
\begin{align}
  V\widetilde{\omega_i}\left[-\frac{b_i}{G_i}+\frac{1}{\widetilde{\omega_i}}\left(k_\omega^{(i)}-\frac{V}{D_{ii}}\right)\right] &= \nonumber \\ 
   V\widetilde{\omega_1}\left[-\frac{b_1}{G_1}+\frac{1}{\widetilde{\omega_1}}\left(k_\omega^{(1)}-\frac{V}{D_{11}}\right)\right],
   \label{equate_ampl_fac}
\end{align}
which leads to,
\begin{equation}
 b_i=\frac{G_i}{\widetilde{\omega_i}}\left[\frac{b_1}{G_1}\widetilde{\omega_1} +
 \left(k_\omega^{(i)}-\frac{V}{D_{ii}}\right) -\left(k_\omega^{(1)}-\frac{V}{D_{11}}\right)\right].
 \label{b_inter_rel}
\end{equation}

Now, using Eq.~\ref{b_inter_rel}, the question of determining $b_i$ corresponding to all the $K-1$ 
components is reduced to the problem of determining $b_1$ only. This is achieved by imposing local 
equilibrium conditions at the perturbed interface, which manifests 
as compositions of the solid and liquid phases calculated from the Gibbs-Thomson condition (refer to the Appendix for details).  

\section{Phase-field model}

The diffuse interface model used to study the current problem of interest is described in ~\cite{Choudhury+11-3}.
The grand-canonical density functional ($\Omega$) of the system is given by:

\begin{align}
 {\Omega}\left(\vmu,T,\phi\right)&=\int_{V}\Bigg[\Psi\left(\vmu,T,
\phi\right) \nonumber \\ 
 &+ \left(\epsilon \sigma|\nabla \phi|^{2} +
\dfrac{1}{\epsilon}w\left(\phi\right)\right)\Bigg]dV.
 \label{GrandPotentialfunctional}
\end{align}

The values of the order parameter ($\phi$, also known as the phase-field) demarcates regions of pure
solid ($\phi=1$), pure liquid ($\phi=0$) and the interface between the two 
(where $\phi$ is a positive fraction), in the solidification microstructure. 
The double-welled polynomial $w(\phi)=9\sigma \phi^{2}\left(1-\phi\right)^{2}$, introduces
a potential barrier between the solid and the liquid phases. 
A penalty in grand potential associated with the gradients in $\phi$ is introduced into the 
model through the term $\sigma |\nabla \phi|^{2}$.    
$\epsilon$ controls the interface width 
and $\sigma$ denotes the interfacial energy respectively.
The grand potential density $\Psi$ of 
the system is obtained by an interpolation of the grand potentials of the solid and the liquid phases,
\begin{align}
 \Psi\left(\vmu,T,\phi\right) &= \Psi^{l}\left(\vmu,T\right)h\left(1-\phi\right)
                                           + \Psi^{s}\left(\vmu,T\right)h\left(\phi\right),
\label{GP_interpolation}
\end{align}
where $\Psi^s$ and $\Psi^l$ are functions of the diffusion potential vector $\vmu=\lbrace{\mu_1,\ldots,\mu_{K-1}\rbrace}$ (assuming a
substitutional alloy under lattice constraint with $K$ components) and the temperature ($T$) in
the system. $h\left(\phi\right) = \phi^{2}\left(3-2\phi\right)$ is an interpolation polynomial with the property 
$h\left(\phi\right) + h\left(1-\phi\right)=1$.

The compositions in every phase can be derived as functions of the diffusion potential vector $\vmu$ as given by:
\begin{align}
 c_{i}^{s,l}\left(\vmu, T\right)=
-V_m\dfrac{\partial \Psi^{s,l}\left(\vmu,T\right)}{\partial \mu_i}.
\label{c_of_mu}
\end{align}
The molar $V_m$ is taken to be a constant across all the components.

Solidification (or melting) is captured by the evolution of $\phi$, which is obtained by solving:
\begin{align}
\tau_{s l} \epsilon \dfrac{\partial \phi}{\partial t}= 2 \sigma \epsilon \nabla^2 \phi-\dfrac{1}{\epsilon}\dfrac{\partial
w\left(\phi\right)}{\partial \phi}-\dfrac{\partial
\Psi\left(\vmu, T, \phi\right)}{\partial \phi}.
\label{Equation6_grandchem}
\end{align}
$\tau_{sl}$ is the relaxation time with its value set to obtain a diffusion controlled interface motion~\cite{Choudhury+12}.

For solidification in a multi-component system, 
the evolution of $\vmu$ can be expressed as:

\begin{align}
 &\left\lbrace\dfrac{\partial \mu_i}{\partial t}\right\rbrace = \nonumber\\
&\left[\sum^{p=s,l} 
h_p\left(\phi\right)\dfrac{\partial c_i^p\left(\vmu,
T\right)}{\partial \mu_j}\right]^{-1}_{ij}\Big\lbrace\nabla\cdot\sum_{j=1}^{K-1}M_{ij}
\left(\phi\right)\nabla \mu_j \nonumber\\ 
&- \sum^{p=s,l} c^p_{i}\left(\vmu,T\right)\dfrac{\partial
h_p\left(\phi\right)}{\partial t}\Big\rbrace.
\label{Mu_explicit_temperature}
\end{align}

where  $\left[\cdot\right]$ denotes 
a matrix of dimension $((K-1) \times (K-1))$  while $\left\lbrace \cdot \right\rbrace$ represents a vector 
of dimension $(K-1)$, and we have used
$h_l\left(\phi\right)=h\left(1-\phi\right)$
and $h_s\left(\phi\right)= h\left(\phi\right)$.
The atomic mobility, $M_{ij}\left(\phi\right)$ is 
obtained by interpolating the individual phase mobilities,
\begin{equation}
 M_{ij}\left( \phi \right) = M_{ij}^l (1-\phi) + M_{ij}^s \phi,
 \label{interp_mobility}
\end{equation}
where, each of the mobility matrix $M_{ij}^{s,l}$ is defined by,
\begin{equation}
 \left[M_{ij}^{s,l}\right] =  \left[D^{s,l}_{ik}\right] \left[\dfrac{\partial
c_k^{s,l}\left(\vmu,T\right)}{\partial \mu_j}\right],
\label{compute_mobility}
\end{equation}
where $D_{ij}^{s,l}$ are the solute inter-diffusivities.

In our calculations, the solute diffusivities are assumed to be negligibly small in the solid compared to the liquid.
This assumption leads to an anomalous solute trapping that is corrected by using an anti-trapping current~\cite{Choudhury+12}.  

\section{Thermodynamics}
The thermodynamic input for the phase-field model comprises of the 
equilibrium compositions of the phases, the inverse of the matrix 
$\left[\dfrac{\partial \mu_i}{\partial c_j}\right]$ which is computed
at the equilibrium compositions of the phases. For computing the matrix
$\left[\dfrac{\partial \mu_i}{\partial c_j}\right]$, in the present work 
we use an ideal solution approximation for representing the free energies, 
 described in detail in the Appendix.
Thereafter, 
we perform a linearization of the phase-diagram 
(mentioned in the Appendix) using the aforementioned
properties giving us the relation between the compositions and the 
chemical potentials as, 
\begin{align}
  \left\lbrace c_{i}^{l,s}\right\rbrace &= \left\lbrace c_{i,eq}^{l,s,*}\right\rbrace 
  +  \left[\dfrac{\partial c_i^{l,s}}{\partial \mu_j}\right]\left\lbrace \mu_j - \mu_{j,eq}^{*}\right\rbrace, 
  \label{c_of_mu_2}
\end{align}
for all the $K-1$ independent components while $\mu_{j,eq}^{*}$ are the values
of the equilibrium diffusion potentials of the components computed at the 
equilibrium compositions $\left(c_{i,eq}^{l,s,*}\right)$, 
about which the properties are linearized.

The driving force for phase transformation ($\Delta \Psi^{ls} (\vmu,T)$) which is 
also required for the phase-field computations is due to 
the difference in grand potentials of the participating phases 
$\Psi^l(\vmu,T)$ and $\Psi^s(\vmu,T)$ which at leading order in 
$\vmu$ reads, 

%
%
\begin{align}
 \Delta\Psi^{ls}=\dfrac{1}{V_m}\left\lbrace c_{i,eq}^{s,*} - c_{i,eq}^{l,*}\right\rbrace \left\lbrace\mu_i-\mu_{i,eq}^{*}\right\rbrace.
 \label{2d_equi}
\end{align}

\section{Results}
In our bid to understand the behavior of $\dot{\delta}/\delta$ as a function of 
$\omega$ ($=2\pi/\lambda$), as predicted by Eq.~\ref{ampl_fac} 
and the phase-field simulations, a Hf-Re-Al-Ni quaternary alloy was selected as the system of study. The equilibrium
tie-line compositions obtained from the Thermotech Ni-
based Superalloys Database (TTNI8) available with the Thermo-Calc software
and the methodology adopted for obtaining the relevant thermodynamic information is mentioned in the Appendix.
While discussing our key observations, we will adhere to the generic representation of solutes by A, B and C, with D
denoting the solvent in order to remain consistent with our theoretical expressions and discussion of 
the phase field modelling technique. Thus, for all practical purposes in the discussion that 
follows, A, B, C and D stand for Hf, Re, Al and Ni respectively.

Henceforth, most of our discussion will revolve around the behavior of the salient features
of a dispersion plot: the maximum in $\dot{\delta}/\delta$ ($(\dot{\delta}/\delta)_{max}$),
the wave number corresponding to the maximum in $\dot{\delta}/\delta$ ($\omega_{max}$), and the
wave number at which $\dot{\delta}/\delta$ changes sign ($\omega_{crit}$), as functions of different parameters.
The significance of $\omega_{max}$ can be understood as the 
parameter which defines the dominating length scale ($\lambda_{max}=2\pi/\omega_{max}$) characterizing 
the early stages of instability, whereas,
$\omega_{crit}$ defines a length scale ($\lambda_{crit}=2\pi/\omega_{crit}$) 
below which the solid-liquid interfacial energy dominates, causing extremely small solidified features to 
melt back. $(\dot{\delta}/\delta)_{max}$ is the term characterizing the magnitude of the fastest growing mode. 
It sets the time scale of the Mullins-Sekerka instability; larger values of which imply 
a quicker destabilization to a cellular or dendritic microstructure.

At the onset, the principal aim of the simulations and the analytical 
studies is to derive the sensitivity of the alloy system towards the 
amplification of the instabilities for different choices of the 
inter-diffusivities of the elements.
Scientifically, this is an 
important question, as these are difficult quantities to determine
experimentally. Secondly, the comparison between the analytical 
calculations and the phase-field simulations also benchmarks the 
model for use in solidification simulations in multicomponent
alloys. Therefore, in this sub-section, we will derive
the dispersion relations using both phase-field simulations
and analytical calculations for a ``given alloy composition"
while we vary the inter-diffusivity matrices.

For setting up the phase-field simulations for comparison to 
the analytical calculations, it is essential to initialize
the system very close to the assumptions that are inherent
in the analysis. Therefore, we start from the phase-field simulation of
a planar growth front in a supersaturated liquid and let the
chemical potential and phase-field profiles develop well into
the scaling regime (where the square of the 
interface position $x_f^{2}$ scales linearly with time with a
defined constant of proportionality $\eta_s$).
Here, the variation of the velocity occurs over
time-scales which are much larger than that required
for the interface to traverse a distance corresponding
to several times the interfacial width. 
The equilibria along with the 
planar front velocity ($V$) recorded from the plane front solidification
simulation, were utilized to compute the composition gradients of different 
components at the interface ($G_i$), all of which appeared in the theoretical
calculation of the dispersion curve. This ensured a uniformity in the parameters 
used across phase field simulations and analytical calculations. 

Thereafter, the 1D composition and phase-field 
profiles in the liquid are perturbed with prescribed wavelengths
defined by the width of the simulation domain possessing 
an amplitude of 5 grid-points which results 
in an initialization where the composition and the phase-field
profiles suitably conform to the imposed morphological
perturbation.

\subsection{Comparison between theory and phase-field simulations}

\begin{figure}[!htbp]
\includegraphics[width=0.9\linewidth]{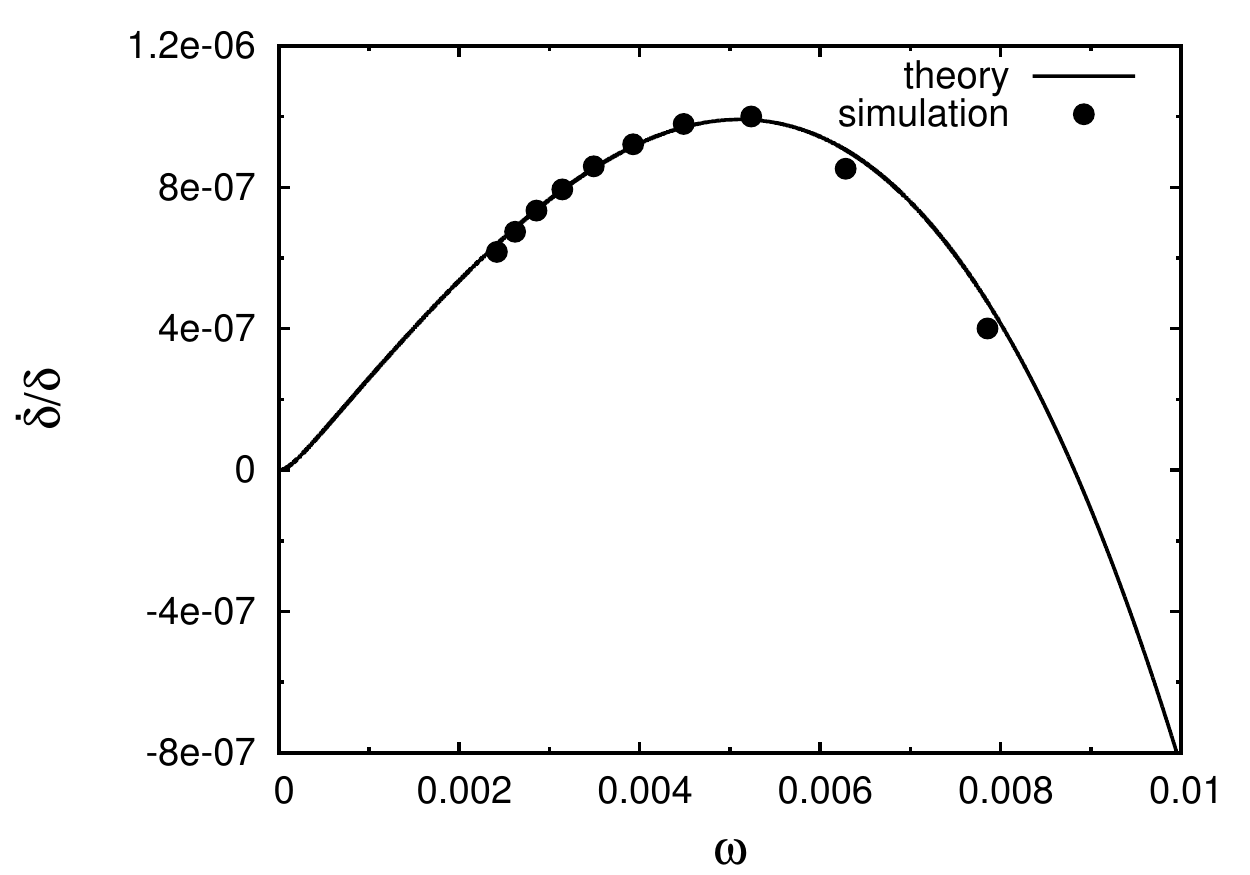}
\caption{$\dot{\delta}/\delta$ versus $\omega$ from phase-field and theoretical calculations. The diffusivity matrix is diagonal with 
$D_{AA}=D_{BB}=D_{CC}=1.0$. The bulk liquid composition chosen for this study is: $c_A=0.001878$, $c_B=0.0165155$, $c_C=0.1871735$.
}
\label{identD}
\end{figure}

We begin our discussion with a comparison of the dispersion plot ($\dot{\delta}/\delta$ versus $\omega$) 
obtained from phase-field simulation against the one predicted by our analytical 
theory for a situation with a diagonal diffusivity matrix with $D_{AA}=D_{BB}=D_{CC}=1.0$ (see Fig.~\ref{identD}).
The graph depicts an excellent agreement between the analysis and simulations\footnote{It must be mentioned at this point that simulations performed for studying the dispersion 
behavior were restricted to modes with $\omega>=0.0024$. This was necessitated by an observed tendency of the system to select 
wavelengths smaller than what the system was initialized with (observed only 
for small values of $\omega$), as we simulated longer in time.}. It must be mentioned here that all calculations (both analytical and phase-field)
are performed with non-dimensionalized parameters whose values and 
the relevant scales required to convert them to dimensional values are mentioned in the Appendix.

\begin{figure}[!htbp]
\includegraphics[width=0.9\linewidth]{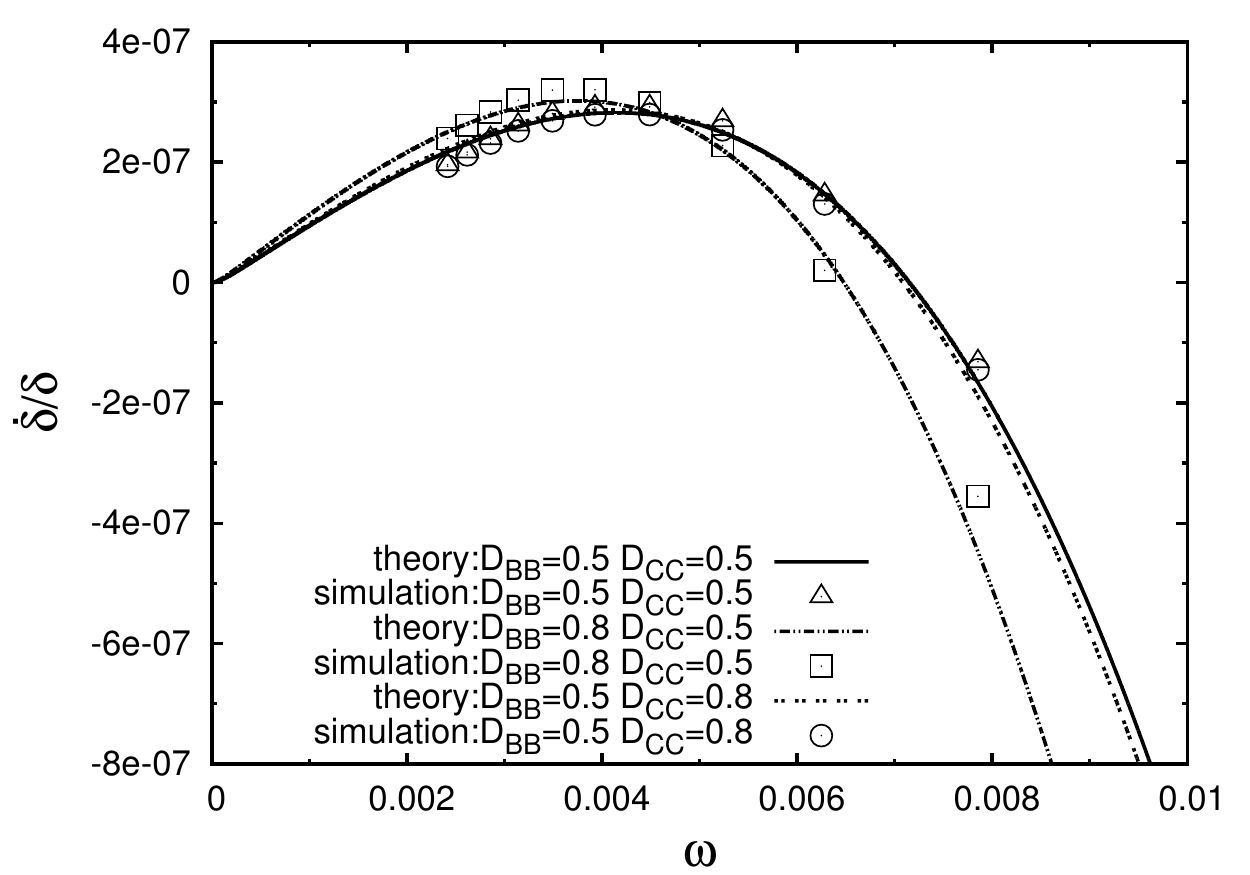}
\caption{$\dot{\delta}/\delta$ versus $\omega$ from phase-field and theoretical calculations. The diffusivity matrix is diagonal with 
$D_{AA}=1.0$ with the other diagonal components being mentioned in the figure legend. The bulk liquid composition chosen here is the same as in 
Fig.~\ref{identD}.}
\label{D_11_10_rest_uneq}
\end{figure}

Moving on to situations where the diffusivity matrix is non-identity.
In Fig.~\ref{D_11_10_rest_uneq}, we present three systems with different combinations 
of $D_{BB}$ and $D_{CC}$ while $D_{AA}$ was held constant.
Here, the analytical curves appear to be more sensitive to a change in $D_{BB}$ 
than to one in $D_{CC}$.

With $D_{BB}=0.8$, the maximally growing wave number ($\omega_{max}$) and the critical 
wave number ($\omega_{crit}$) move towards left with a slight increase 
in the magnitude of $(\dot{\delta}/\delta)_{max}$ compared to the cases where $D_{BB}=0.5$. 
Thus, it appears that a higher value of $D_{BB}$ leads to the selection of larger length scales of instability. 
The dispersion plot for $D_{BB}=0.8$ also appears to have a different shape than the ones with $D_{BB}=0.5$, 
 which can be discerned by the lower growth rates for wave numbers larger than $\omega_{max}$ in the 
former than in the latter, with the reverse being true for 
$\omega<\omega_{max}$. This observed change in shape is a natural
outcome of the changes in $\omega_{max}$, $\omega_{crit}$ and 
$(\dot{\delta}/\delta)_{max}$ effected by a change in $D_{BB}$.

\begin{figure}[!htbp]
\includegraphics[width=0.9\linewidth]{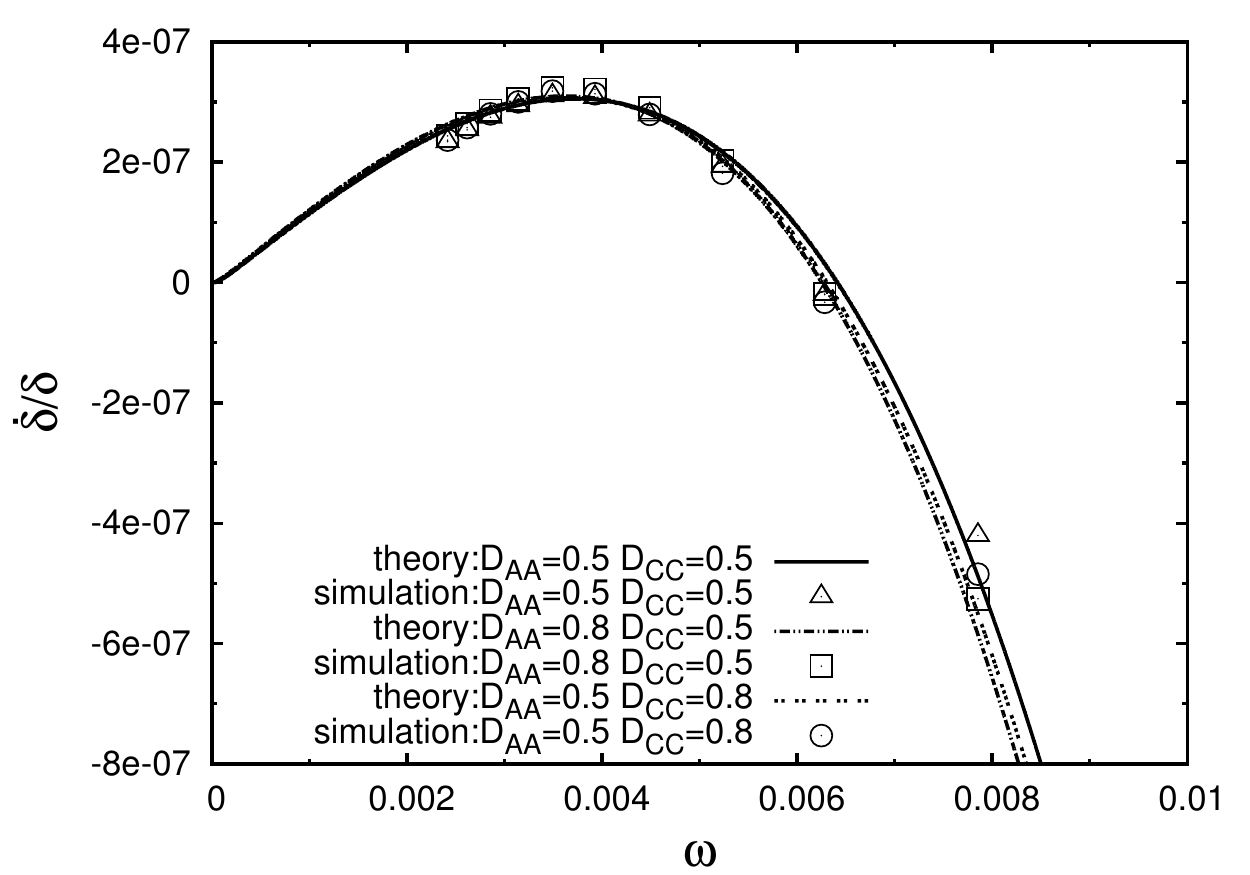}
\caption{$\dot{\delta}/\delta$ versus $\omega$ from phase-field and theoretical calculations. The diffusivity matrix is diagonal with 
$D_{BB}=1.0$ with the other diagonal components being mentioned in the figure legend. The bulk liquid composition chosen here is the same as in 
Fig.~\ref{identD}. }
\label{D_22_10_rest_uneq}
\end{figure}

In Fig.~\ref{D_22_10_rest_uneq}, we study the dispersion relations by maintaining 
$D_{BB}=1.0$ for three systems with different 
combinations of $D_{AA}$ and $D_{CC}$ (lower than $D_{BB}$). The plots appear to superimpose 
upholding the observation in Fig.~\ref{D_11_10_rest_uneq} that the dispersion relations 
are a lot more sensitive to changes in $D_{BB}$ than to modifications
in $D_{AA}$ and $D_{CC}$. Thus, despite variations in $D_{AA}$ and $D_{CC}$, 
the instability dynamics and the relevant critical 
wavelengths remain largely invariant due to constancy of $D_{BB}$.

\begin{figure}[!htbp]
\includegraphics[width=0.9\linewidth]{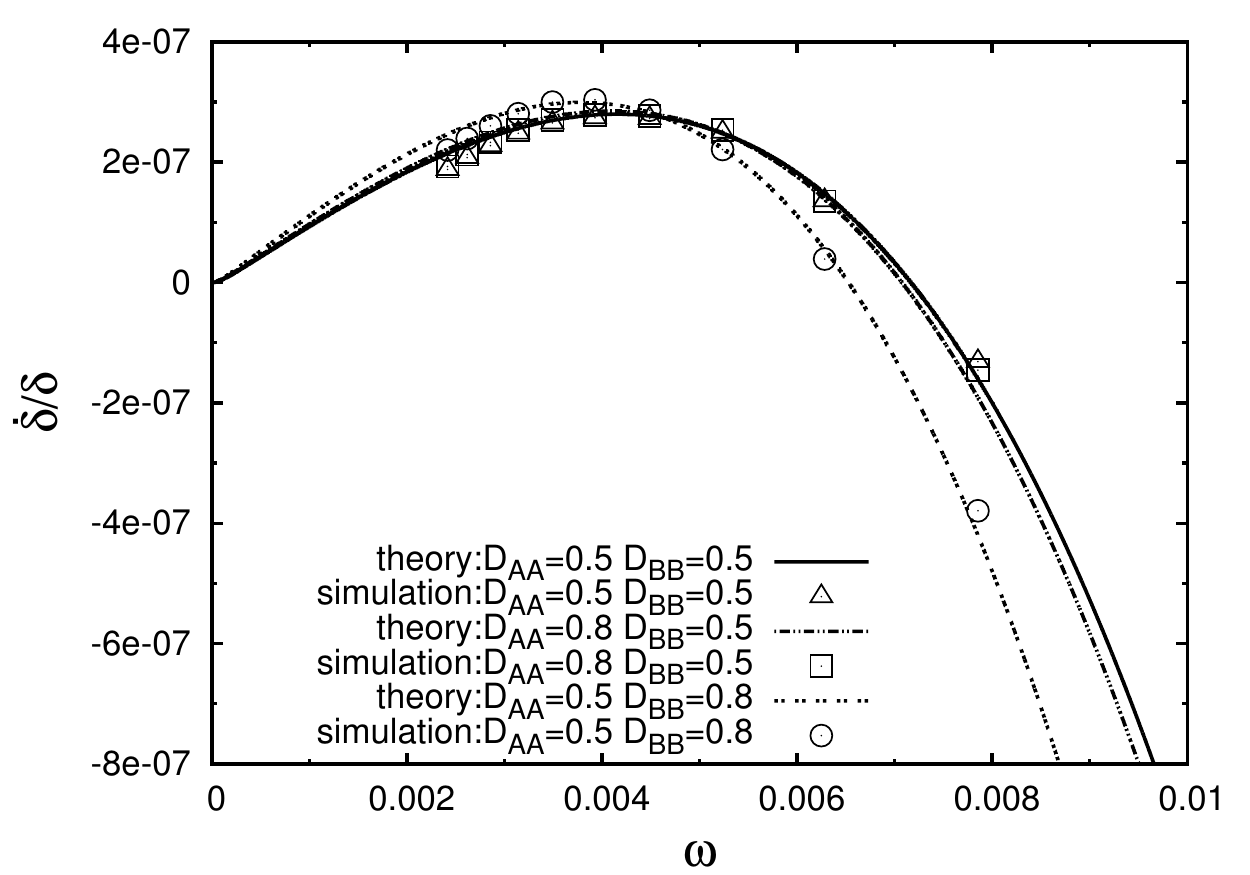}
\caption{$\dot{\delta}/\delta$ versus $\omega$ from phase-field and theoretical calculations. The diffusivity matrix is diagonal with 
$D_{CC}=1.0$ with the other diagonal components being mentioned in the figure legend. The bulk liquid composition chosen here is the same as in 
Fig.~\ref{identD}. }
\label{D_33_10_rest_uneq}
\end{figure}

Subsequently, we investigate the situation where we fix $D_{CC}=1.0$ and vary $D_{AA}$ and $D_{BB}$ 
(as shown in Fig.~\ref{D_33_10_rest_uneq}), for which trends very similar to what observed in Fig.~\ref{D_11_10_rest_uneq}, 
is retrieved. A higher value of $D_{BB}$ resulted in smaller
values for $\omega_{max}$ and $\omega_{crit}$, 
with the wave numbers smaller
than $\omega_{max}$  growing faster than in systems with a lower value of $D_{BB}$ and vice-versa. 
Thus, the consequent microstructural implications derivable from Fig.~\ref{D_33_10_rest_uneq}, 
are similar to what was discussed for 
Fig.~\ref{D_11_10_rest_uneq}. 
Also, as observed in Fig.~\ref{D_11_10_rest_uneq}, 
a higher value in $D_{BB}$ resulted in a higher value for $(\dot{\delta}/\delta)_{max}$.

The preceding results established the major trends in the variation of the dispersion relations as functions of diffusivity. It must
be reiterated at this point that though the explicit changes were only in the diffusivity matrices, there were also associated
changes in the equilibrium tie-line compositions of the planar profiles along with changes in velocity of the interface.
So in the studies discussed till now, the inferences we draw upon can be seen as a result of combined changes in the equilibrium
compositions at the interface, the velocity and the diffusivities (the equilibrium tie-line compositions and the velocities
are enumerated in Table~\ref{tie_lines_pfm}. 

\begin{table*}[!htbp]
\centering
\begin{tabular}{|c|c|c|c|c|c|c|c|c|c|}
\hline
 $D_{AA}$ & $D_{BB}$ & $D_{CC}$ & $V$ & $c_{A,eq}^s$ & $c_{B,eq}^s$ & $c_{C,eq}^s$ &  $c_{A,eq}^l$ & $c_{B,eq}^l$ & $c_{C,eq}^l$\\ 
 \hline
 1.0     &0.5  	 &0.5    &1.07e-04       &0.000517       &0.023554       &0.177660       &0.002966       &0.009170       &0.197032 \\  
 \hline
 1.0	 &0.8	 &0.5	 &1.26e-04	 &0.000539	 &0.023891	 &0.178480	 &0.003092	 &0.009282	 &0.197929 \\  
 \hline
 1.0	 &0.5	 &0.8	 &1.10e-04	 &0.000520	 &0.023404	 &0.176777	 &0.002983	 &0.009112	 &0.196085 \\
 \hline
 0.5	 &1.0	 &0.5	 &1.28e-04	 &0.000600	 &0.024395	 &0.178584	 &0.003437	 &0.009469	 &0.198042 \\
 \hline
 0.5	 &1.0	 &0.8	 &1.31e-04	 &0.000606	 &0.024246	 &0.177609	 &0.003467	 &0.009411	 &0.196998 \\
 \hline
 0.8	 &1.0	 &0.5	 &1.33e-04	 &0.000564	 &0.024154	 &0.178789	 &0.003237	 &0.009377	 &0.198267 \\
 \hline
 0.5	 &0.5	 &1.0	 &1.06e-04	 &0.000562	 &0.023626	 &0.176222	 &0.003219	 &0.009198	 &0.195484 \\
 \hline
 0.5	 &0.8	 &1.0	 &1.24e-04	 &0.000592	 &0.023997	 &0.176873	 &0.003392	 &0.009323	 &0.196198 \\
 \hline
 0.8	 &0.5	 &1.0	 &1.10e-04	 &0.000533	 &0.023421	 &0.176352	 &0.003056	 &0.009121	 &0.195626 \\
 \hline
\end{tabular}
\caption{Tie-line compositions and steady-state velocities as functions of diffusivities.}
\label{tie_lines_pfm}
\end{table*}

Thus, to isolate and understand the influence of each of the contributing factors better, we 
resort to more controlled studies, where we vary only one of the parameters at a time while maintaining the others 
constant. They are sequentially described in the following subsections. 

\subsection{Influence of diffusivity}
In the first among these, we will 
investigate the influence of the change in the diffusivity matrices keeping both 
the velocity as well as the tie-line compositions fixed.In a separate work~\cite{Lahiri2016_1},
we study how the equilibrium compositions at the interface can 
be determined for different choices of diffusivity matrices. Drawing upon this, 
we isolate alloy compositions giving us the same growth coefficient ($\eta_s
=x_f/\sqrt{t}$, $x_f$ being the position of the solid-liquid interface at a time $t$)
for different values of the diffusivity matrices, ensuring at all times that the equilibrium
compositions at the interface remain invariant (at the chosen thermodynamic tie-line). 
For each of these conditions, we can then 
derive the characteristics of the dispersion behavior which are highlighted
by the quantities $(\dot{\delta}/\delta)_{max}$, $\omega_{max}$ and $\omega_{crit}$
derived using the analytical expressions described in the previous 
sections, where the velocity $V$ ($=\eta_s/\sqrt{t}$) is computed using an arbitrary time $t$, which 
is kept the same for all the diffusivity combinations. 
\begin{figure}[!htbp]
\centering
\subfigure[]{\includegraphics[width=0.8\linewidth]{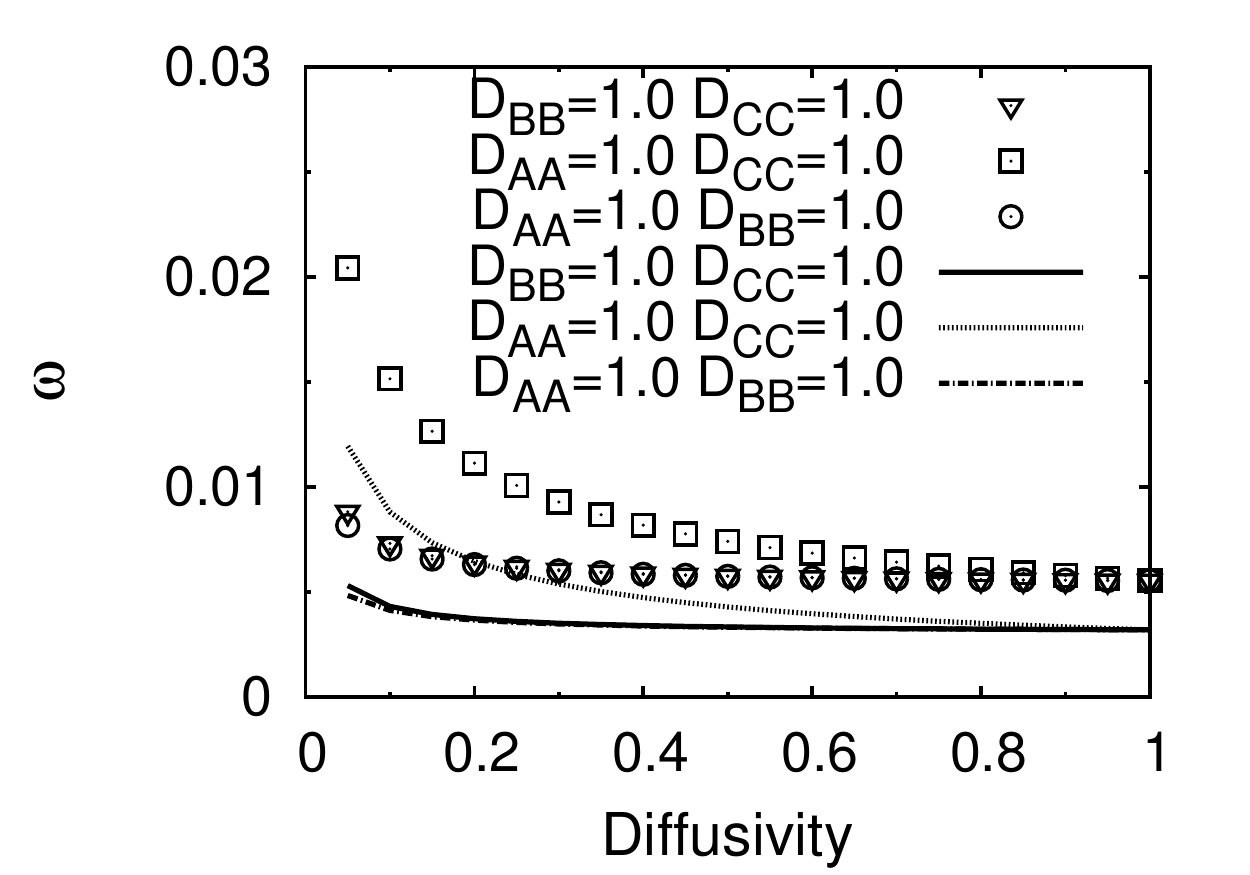}
\label{omega_vs_D}
}
\subfigure[]{\includegraphics[width=0.8\linewidth]{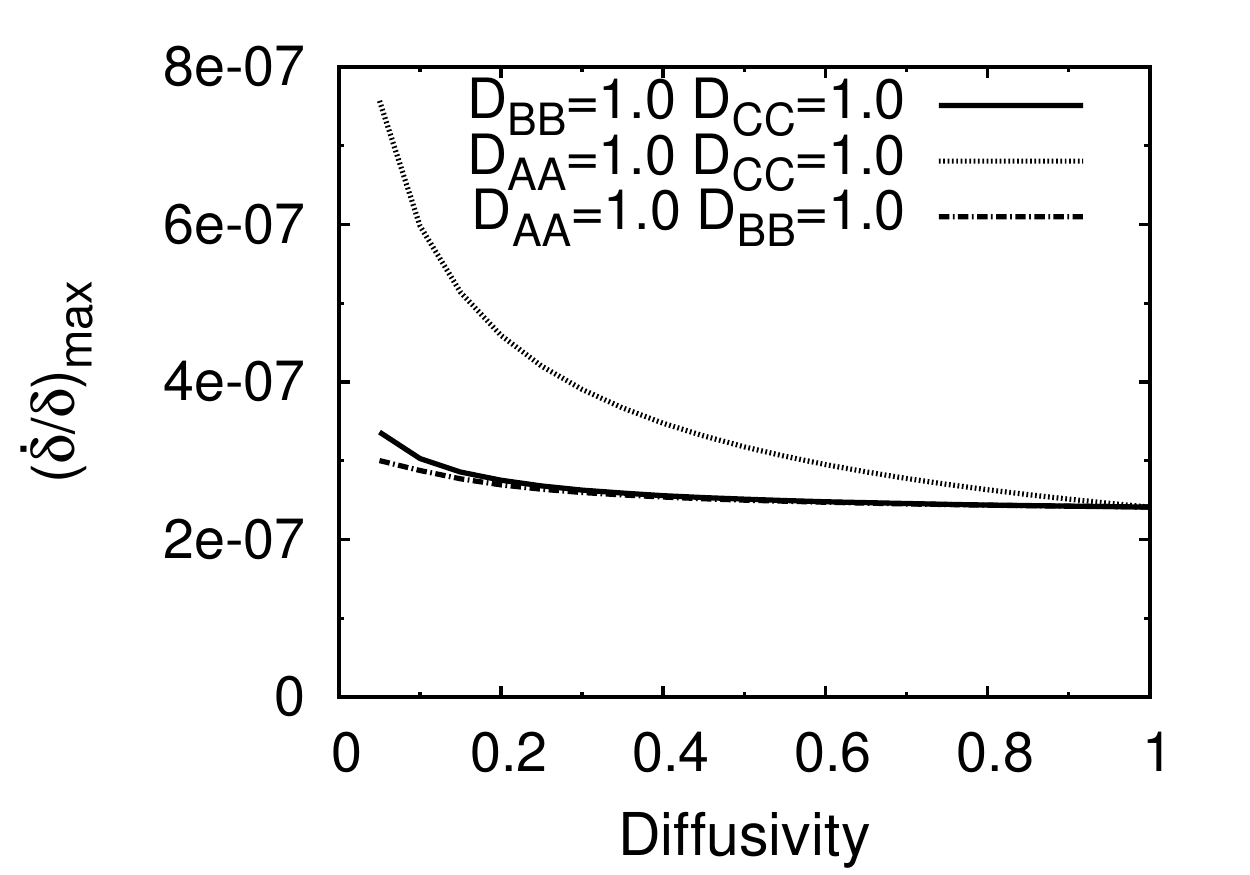}
\label{ampl_max_vs_D}
}
\caption{Plots showing variations of, (a)$\omega_{max}$ (shown by lines) and $\omega_{crit}$ (shown by points), (b) $(\dot{\delta}/\delta)_{max}$,
with change in diffusivity of any one of the components
$A$, $B$ and $C$ (the diffusivity matrix was diagonal).
The figure legends reveal the components whose diffusivities are held constant along with the values set for them, while the 
 diffusivity of the component not mentioned in the legend is varied to generate the curve. The value of $\eta_s$ was selected to be $0.33$.
 The time chosen for analysis was $t=2000000$.}
\label{var_with_D}
\end{figure}

In Fig.~\ref{omega_vs_D}, we report the variation of $\omega_{max}$ and $\omega_{crit}$ 
for three different cases, where in each of the situations either of 
$D_{AA}$, $D_{BB}$, $D_{CC}$ is varied while the others are left constant. 
We see that the $\omega_{max}$ (also $\omega_{crit}$) decreases smoothly with increase in diffusivity
for all the three cases; the pronounced effect (steeper sensitivity to changes
in diffusivity) being for the case where 
only $D_{BB}$ is varied while $D_{AA}$ and $D_{CC}$ are held constant.
The observations are consistent with the fact that
as the interdiffusivities of individual components ($D_{AA}$, $D_{BB}$ or $D_{CC}$) 
increases while $V$ and the equilibrium tie-line compositions stay the same, the composition gradients 
at the planar interface ($G_A$,$G_B$ and $G_C$) decreases. This decrease can be qualitatively
linked to the weakening of the diffusive instabilities which shifts the critical 
wavelengths to larger values (or reduction of $\omega_{max}$ and $\omega_{crit}$)
as highlighted in from Fig.~\ref{omega_vs_D}.

The interesting fact is the difference in the sensitivities
towards change in the inter-diffusivities. For $D_{BB}$ ($<1$) the system chooses much larger 
values for $\omega_{max}$ and $\omega_{crit}$ as compared to the values chosen for 
the same magnitudes of $D_{AA}$ and $D_{CC}$. This behavior can be explained on the 
basis of the changes in the diffusion length scales (brought about by the 
change in the diffusivities) and capillary length scales described
principally by the Gibbs-Thomson coefficients (and therefore the equilibrium compositions).

In a binary alloy, the changes in the critical length scales, 
denoted by ($2\pi/\omega_{crit}$) and ($2\pi/\omega_{max}$), 
approximately scale with $\sqrt{d_ol_D}$, where $d_o$ is the capillary length, 
and $l_D$ is the diffusion length, each of which can be controlled independently.
Thereby, a binary alloy with larger capillary length $d_o$, will show a higher
sensitivity towards changes in $l_D$ with relation to their influence on the critical 
length scales ($2\pi/\omega_{crit}$) and ($2\pi/\omega_{max}$).

However, in ternary and higher multi-component alloys this correlation becomes difficult to 
establish as the influence of capillarity (principally related to the term $b_i$ (see Eq.~\ref{pert_interf})) 
becomes coupled to the changes in diffusivity (see Eq.~\ref{b_inter_rel}). 
Therefore, the material properties which govern 
the value $b_i$ (which is principally just related to the Gibbs-Thomson coefficient in the 
case of binary alloys) cannot be explicitly linked to the thermodynamic properties
of the phases $\left[\dfrac{\partial \mu_i}{\partial c_j}\right]$.
In essence, this implies that changes in the capillary lengths and
the diffusion lengths become inter-related.

\begin{figure}[!htbp]
\centering
\subfigure[]{\includegraphics[width=0.8\linewidth]{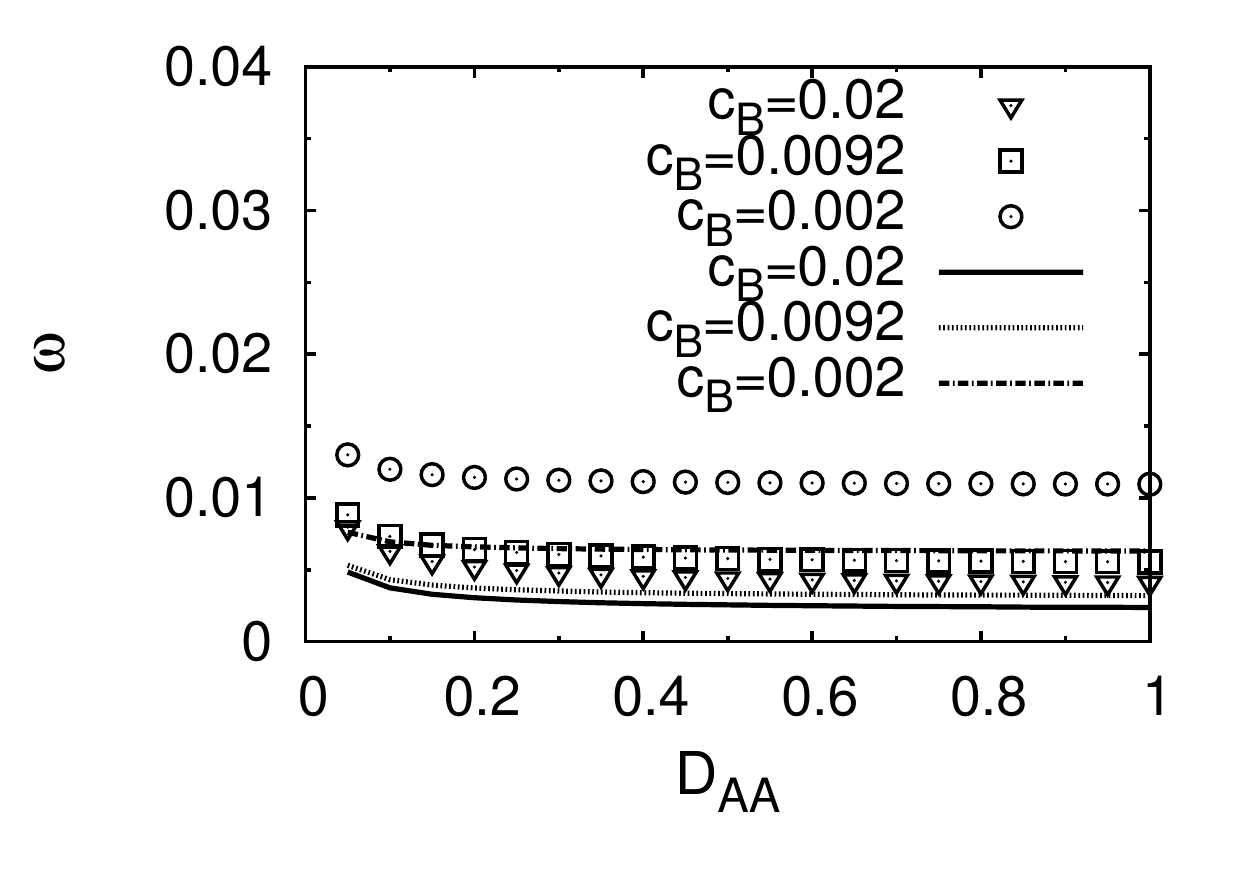}
\label{omega_vs_D_AA}
}
\subfigure[]{\includegraphics[width=0.8\linewidth]{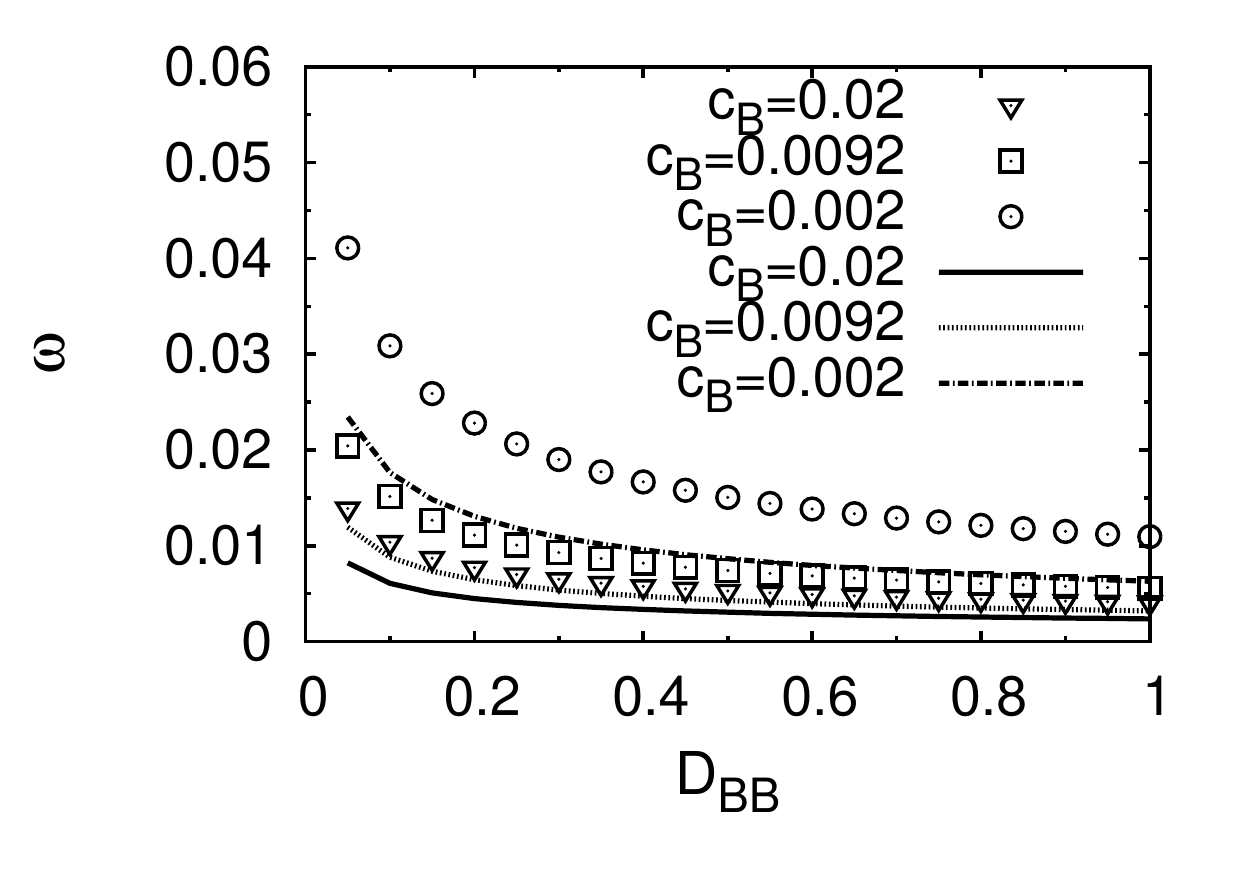}
\label{omega_vs_D_BB}
}
\subfigure[]{\includegraphics[width=0.8\linewidth]{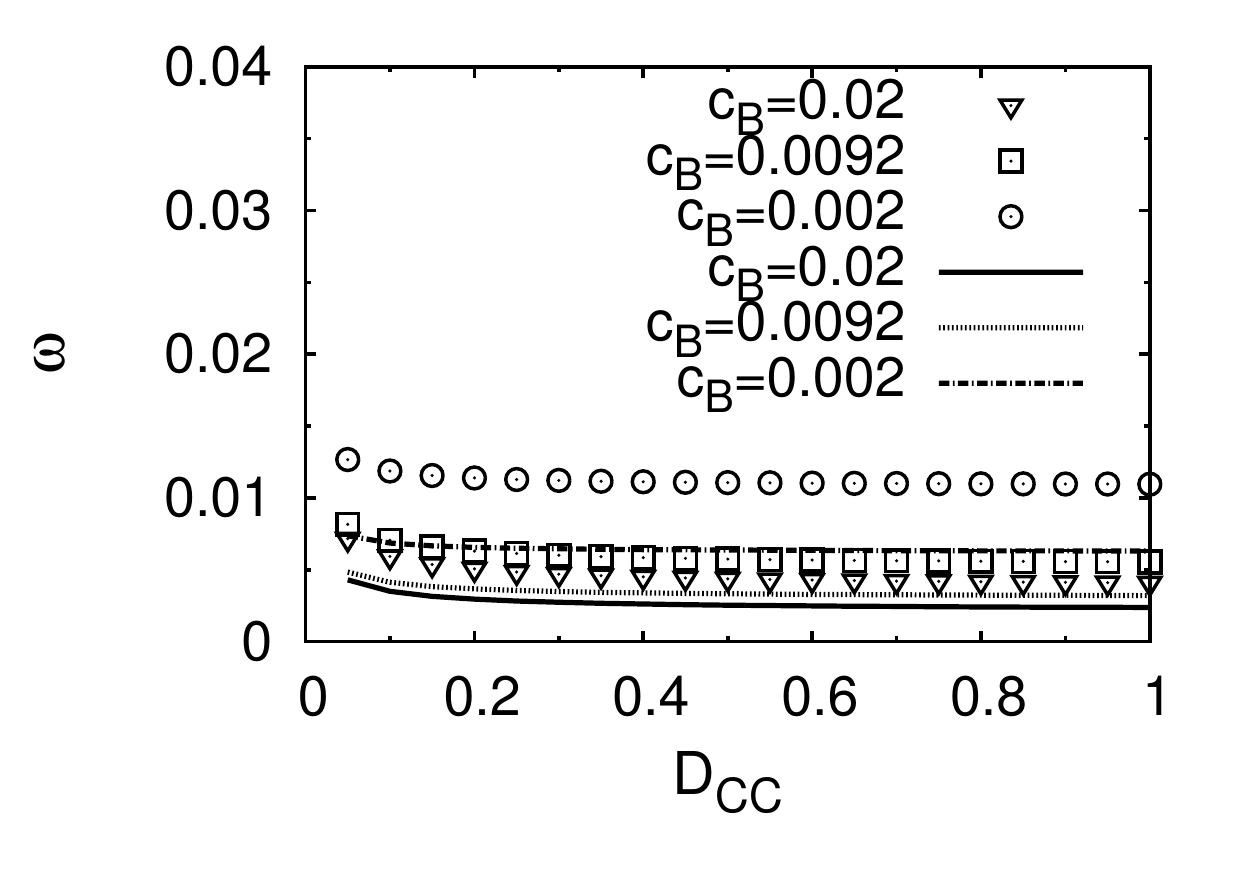}
\label{omega_vs_D_CC}
}
\caption{Plots showing variations of $\omega_{max}$ (shown by lines) and $\omega_{crit}$ (shown by points)
with change in diffusivity of the component
A in (a), B in (b) and C in (c) each for different values of the composition $c_B$ (the diffusivity of the other 
two components were held at unity).}
\label{var_with_D_C}
\end{figure}

In order to establish this inter-dependence we numerically evaluate the length scales for 
different equilibrium 
compositions by changing arbitrarily the composition $c_B$ (which essentially
changes the $\left[\dfrac{\partial \mu_i}{\partial c_j}\right]$ according to Eq.~\ref{gen_for_dc_dmu} 
in Appendix) of both phases maintaining
the same partitioning between the phases ($\Delta c_{B,eq}=c_{B,eq}^l-c_{B,eq}^s$, remaining constant)
and the results are plotted in Fig.\ref{var_with_D_C}.

We see that with lower compositions of $c_B$ the changes in the critical  
length scales become more sensitive to the changes of the diffusivity for all
three components with the maximum sensitivity being still for the 
changes in diffusivity of element B. 

Further, the consequence of the value of segregation can also be coupled
with the above discussion to qualitatively interpret the results displayed
in Fig.\ref{var_with_D}. Here, the amount of segregation for the component
A is the least while we have similar magnitudes of 
segregation for the element B and C. So, we can qualitatively conclude, 
that a smaller phase composition in combination with larger segregation leads to 
greater sensitivity towards the change in length scales as the diffusivity
of that particular component is varied.

The variation in $(\dot{\delta}/\delta)_{max}$ with different component diffusivities show a trend in Fig.~\ref{ampl_max_vs_D}
which can be explained by considering Eq.~\ref{ampl_fac} and the fact that $\widetilde{\omega_B}$ has a 
magnitude which is much higher than that of $\widetilde{\omega_A}$ and $\widetilde{\omega_C}$ at $\omega_{max}$'s corresponding 
to respective variations in $D_{BB}$, $D_{AA}$ and $D_{CC}$. This in turn is a result of a combination of larger compositional 
gradients and lower values of compositions (since $\tilde{\omega}_i = k_\omega^{(i)} + G_i/c_i$).

\subsection{Influence of velocity}
\begin{figure}[!htbp]
\centering
\subfigure[]{\includegraphics[width=0.8\linewidth]{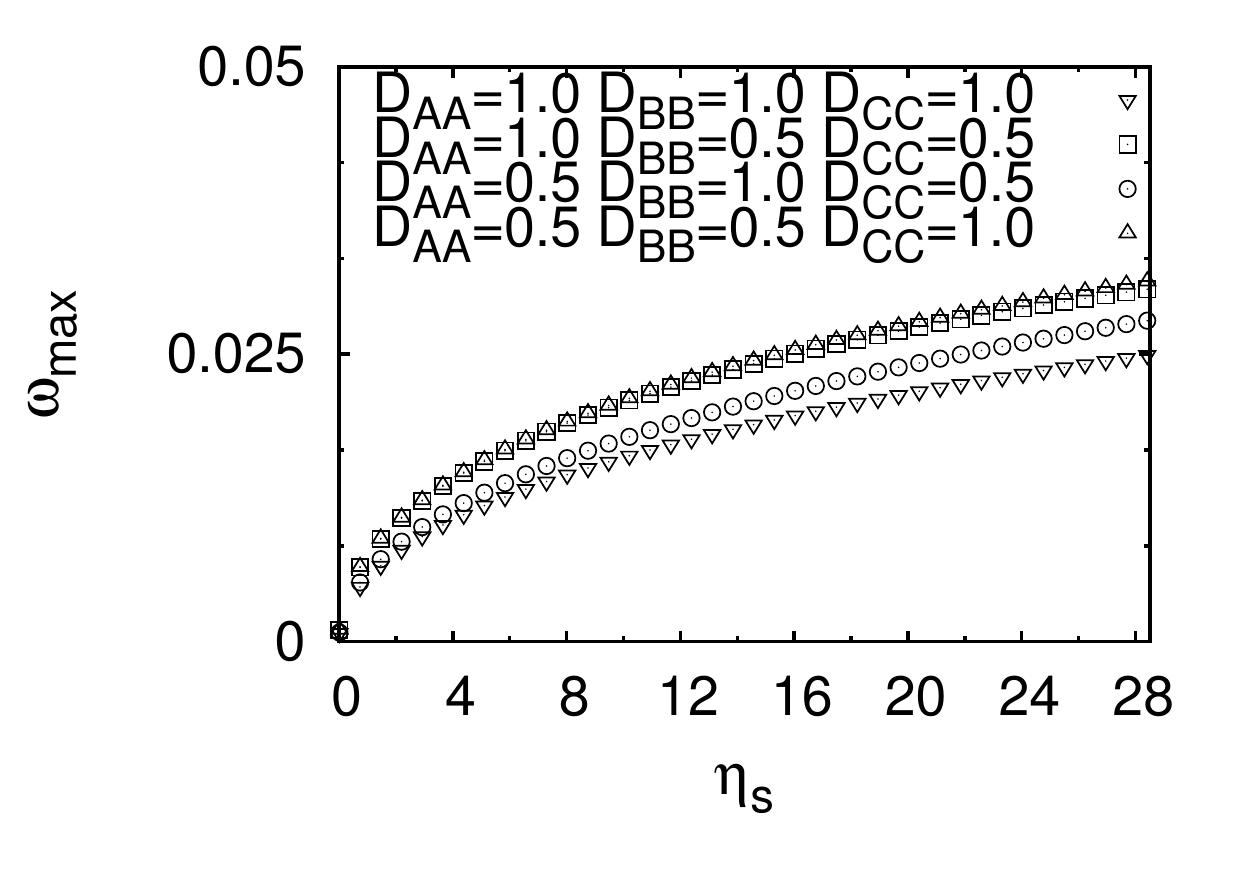}
\label{omega_max_vs_eta_s}
}
\subfigure[]{\includegraphics[width=0.8\linewidth]{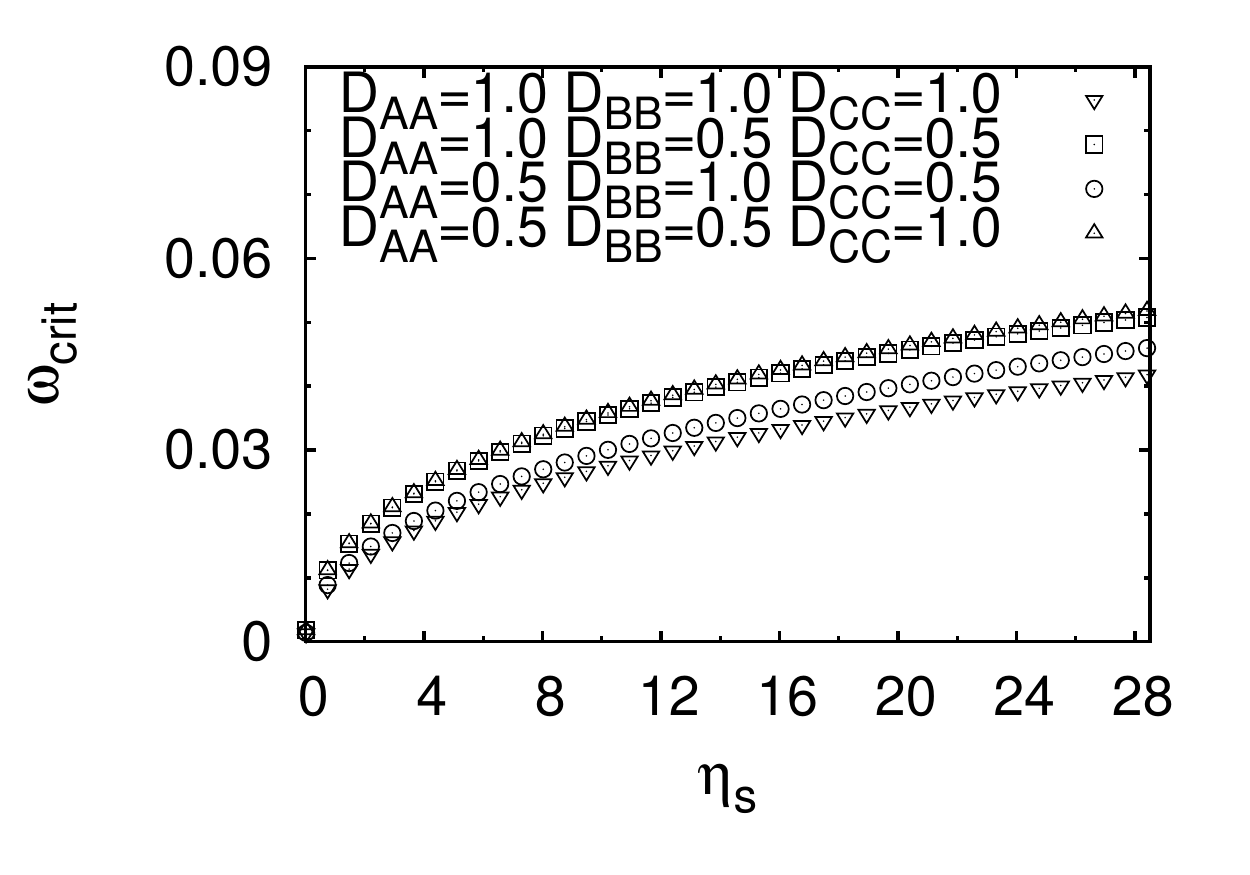}
\label{omega_crit_vs_eta_s}
}
\subfigure[]{\includegraphics[width=0.8\linewidth]{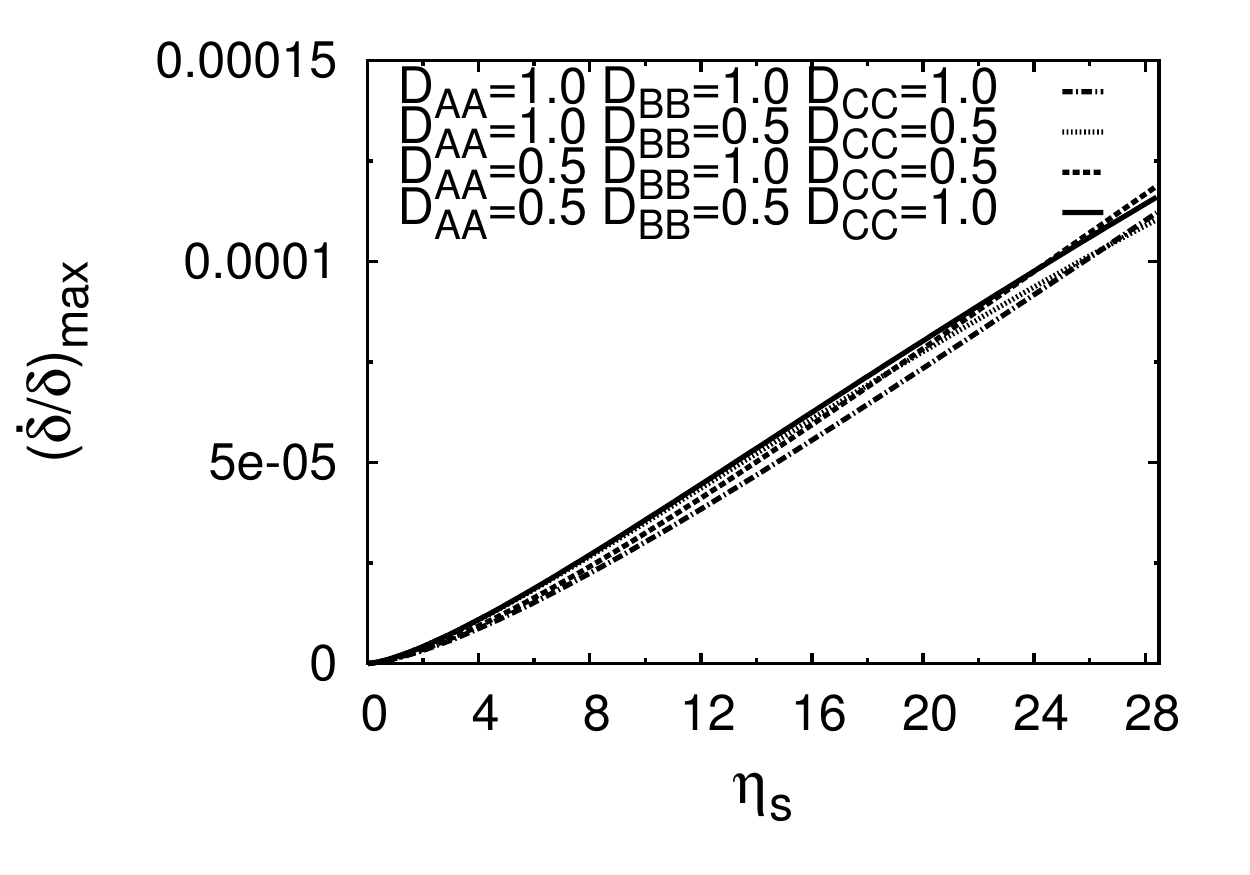}
\label{ampl_max_vs_eta_s}
}
\caption{Plots showing variations of (a)$\omega_{max}$, (b) $\omega_{crit}$, (c) $(\dot{\delta}/\delta)_{max}$,
with $\eta_s$ for different diagonal diffusivity matrices. The time chosen for analysis was $t=2000000$.}
\label{var_with_eta_s}
\end{figure}

Recalling from our work~\cite{Lahiri2016_1} that a multitude of bulk alloy compositions, each corresponding 
to a different $\eta_s$, can grow with the same equilibrium tie-line
compositions (mentioned in the Appendix) for a given diffusivity matrix, allows us to investigate the effects of $\eta_s$ 
(which essentially controls the plane front interface velocity, $V$)
on parameters such as $\omega_{max}$, $\omega_{crit}$ and $(\dot{\delta}/\delta)_{max}$. 
We begin our discussion with Figs.~\ref{omega_max_vs_eta_s} and~\ref{omega_crit_vs_eta_s} 
where $\omega_{max}$ and $\omega_{crit}$ are plotted
against $\eta_s$. The general trend of the curves point towards a selection of 
larger values of $\omega_{max}$ and $\omega_{crit}$ for higher values of $\eta_s$. 
This is due to the fact that higher planar front velocities ($V$) associated with higher values 
of $\eta_s$ leads to higher magnitudes of composition gradients at the planar interface 
$G_i$, which translates to smaller diffusion lengths and consequently the critical and 
maximally growing wavelengths. The curves corresponding to $D_{BB}=1.0$ show a 
much gentler increase with $\eta_s$ compared to the ones with $D_{BB}=0.5$, 
which can be explained based on the discussion accompanying
Fig.~\ref{omega_vs_D}.
The increase in planar front velocity ($V$) with $\eta_s$ is the reason behind 
the consequent increase in $(\dot{\delta}/\delta)_{max}$ 
(see Fig.~\ref{ampl_max_vs_eta_s}), 
as can be seen from Eq.~\ref{ampl_fac}. 
Furthermore, as discussed earlier, a higher $V$ also leads to a higher 
$G_i$, which further increases the value of $(\dot{\delta}/\delta)_{max}$ 
(see first term in brackets of Eq. \ref{ampl_fac} ($b_i/G_i$) which 
decreases with increase in $G_i$).

\subsection{Influence of alloy compositions on the thermodynamic tie-line}
Having individually understood the influence of the velocities and the diffusivities on the Mullins-Sekerka
instability, we can now attempt to understand another particular question of engineering 
importance, which is the prediction of $\omega_{max}$, $\omega_{crit}$ and $(\dot{\delta}/\delta)_{max}$ 
for different alloy compositions along a given thermodynamic tie-line. 
This study incorporates the changes in the behavior of perturbations 
owing to the coupled changes in the velocity and the equilibrium tie-lines
that are selected. Therefore, this 
requires a knowledge of the tie-lines that will be selected by the system during 
planar front growth and the value of $\eta_s$ characterizing that steady state regime,
which can be computed from the expressions mentioned in the Appendix~\cite{Lahiri2016_1}. 
We perform this study on a series of bulk compositions lying along a pre-selected tie-line (mentioned in the Appendix) 
resulting in certain interesting trends worth examining.

To this end, we define a parameter $\nu$, which denotes
the volume fraction of the solid; a smooth variation in which, allows consideration of different 
bulk compositions given by: $c_{i,eq}^s\nu+c_{i,eq}^l(1-\nu)$.   

\begin{figure}[!htbp]
\centering
\subfigure[]{\includegraphics[width=0.8\linewidth]{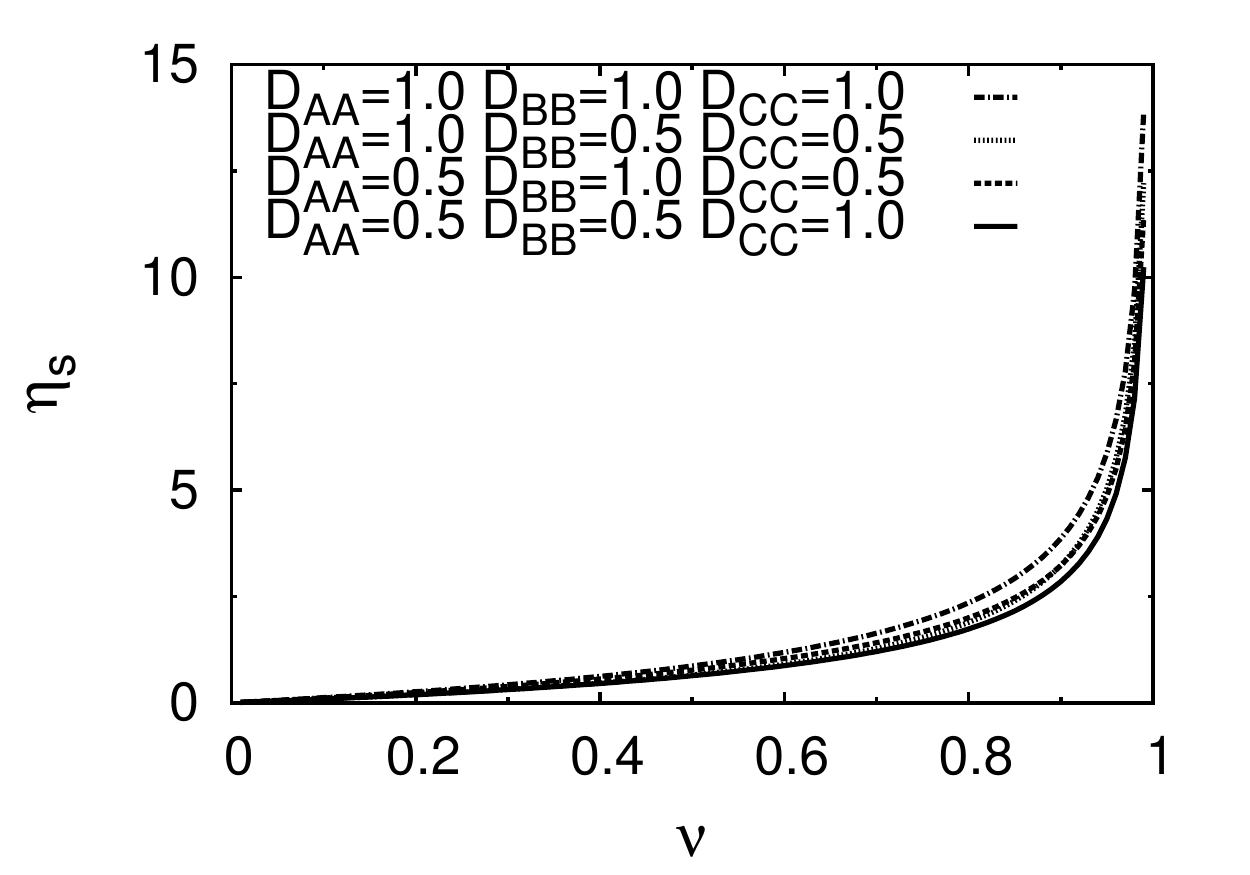}
\label{eta_s_vs_nu}
}
\subfigure[]{\includegraphics[width=0.8\linewidth]{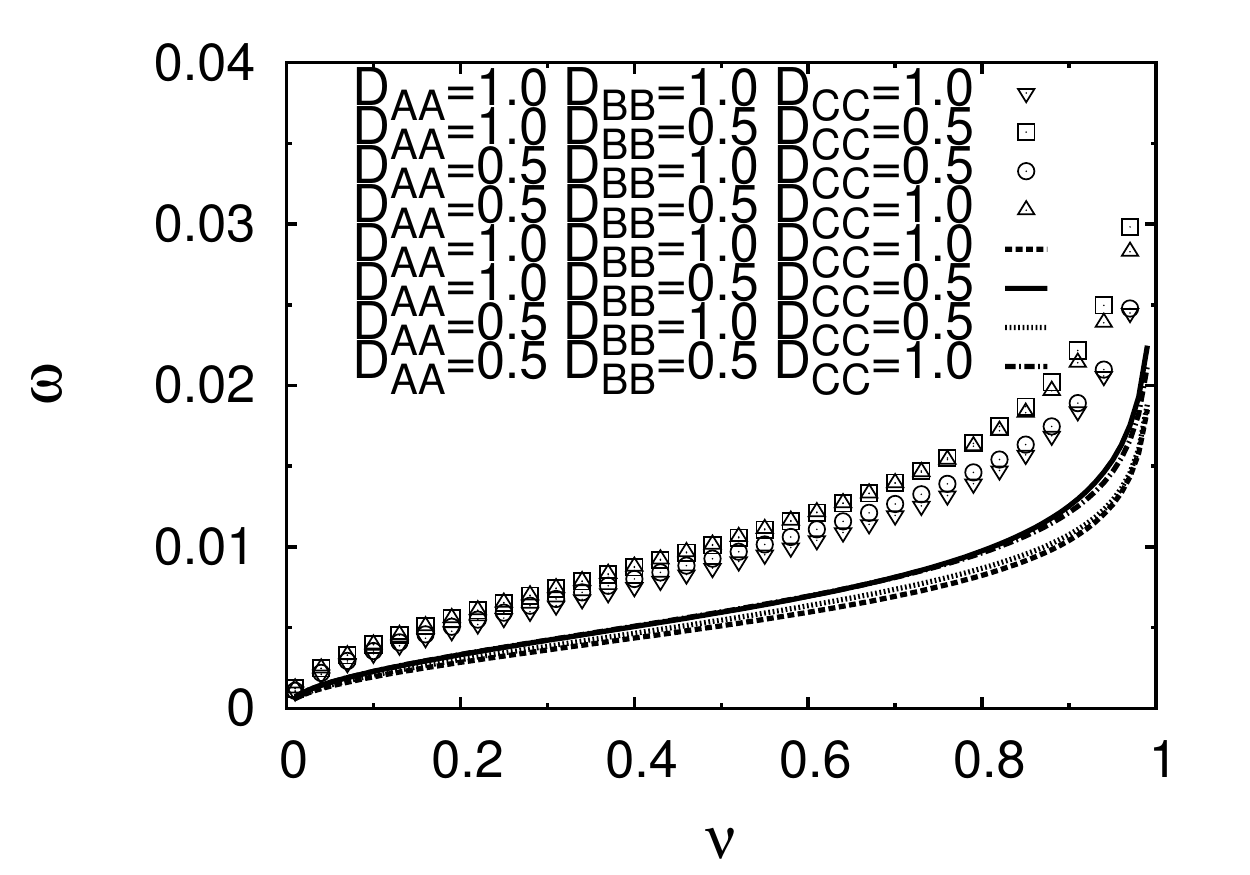}
\label{omega_vs_nu}
}
\subfigure[]{\includegraphics[width=0.8\linewidth]{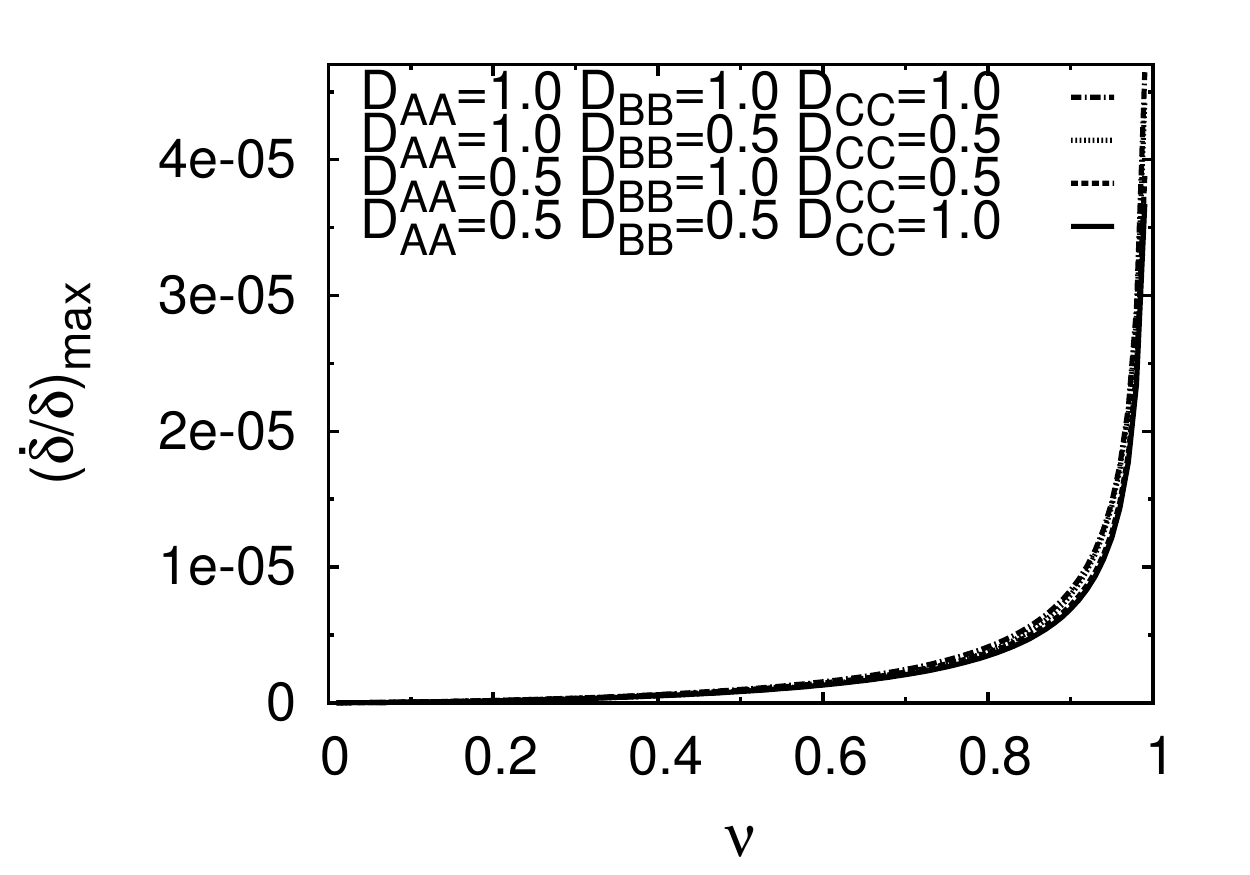}
\label{ampl_max_vs_nu}
}
\centering
\caption{Plots showing variations of (a)$\eta_s$, (b)$\omega_{max}$ (shown by lines) and 
$\omega_{crit}$ (shown by points), (c) $(\dot{\delta}/\delta)_{max}$,
with $\nu$ for different diagonal diffusivity matrices. The time chosen for analysis was $t=2000000$.}
\label{var_with_nu}
\end{figure}

From Fig.~\ref{eta_s_vs_nu}, we can see that $\eta_s$ increases with increasing $\nu$. This is an important information 
as $\eta_s$ sets the velocity ($V$) of the planar interface, which in turn affects 
the composition gradients at the steady-state interface ($G_i$) and higher values of 
both of these parameters have a propensity to make the
system more susceptible to growth of perturbations. It must be mentioned here that as we choose 
different alloy compositions along a given thermodynamic tie-line, the continuous variation in the
selection of steady-state growth velocities is accompanied by the system equilibrating at different 
thermodynamic tie-lines. Thus, different $\Delta c_i=c_{i,eq}^l-c_{i,eq}^s$'s are chosen for 
different bulk alloy compositions, which also serve to modify the composition gradients at the interface
during steady-state growth.

In light of the information presented in Fig.~\ref{eta_s_vs_nu}, we 
can attempt to understand the increase of $\omega_{max}$ and 
$\omega_{crit}$ with $\eta_s$
as shown in Fig.~\ref{omega_vs_nu}. This can be explained by the higher 
$V$ and $G_i$'s associated with higher values of $\eta_s$, which 
causes the system to choose smaller length scales manifesting as higher
values of $\omega_{max}$ and $\omega_{crit}$. 
The reasons for curves corresponding to $D_{BB}=0.5$ 
reporting higher values than ones with $D_{BB}=1$ have 
already been discussed in conjunction with Fig.~\ref{omega_vs_D}.

From Fig.~\ref{ampl_max_vs_nu} we can find $(\dot{\delta}/\delta)_{max}$ varying against $\nu$ in a manner similar to what 
observed for Fig.~\ref{ampl_max_vs_eta_s}. The increase in $\eta_s$ with $\nu$ as shown in Fig.~\ref{eta_s_vs_nu}, explains the
increase in $(\dot{\delta}/\delta)_{max}$ using the arguments associated with Fig.~\ref{ampl_max_vs_eta_s}. 

\subsection{Influence of off-diagonal terms in the diffusivity matrix}
For systems where there is a significant coupling between the diffusion of different solutes, 
it gives rise to off-diagonal terms in the
diffusivity matrix. This introduces complications which preclude most of the analytical 
techniques we have successfully employed till now.
Therefore, we utilize numerical phase-field simulations as a method of studying 
the Mullins-Sekerka type instabilities, where again 
we examine the variation of $(\dot{\delta}/\delta)$ with $\omega$.
\begin{figure}[!htbp]
\includegraphics[width=0.9\linewidth]{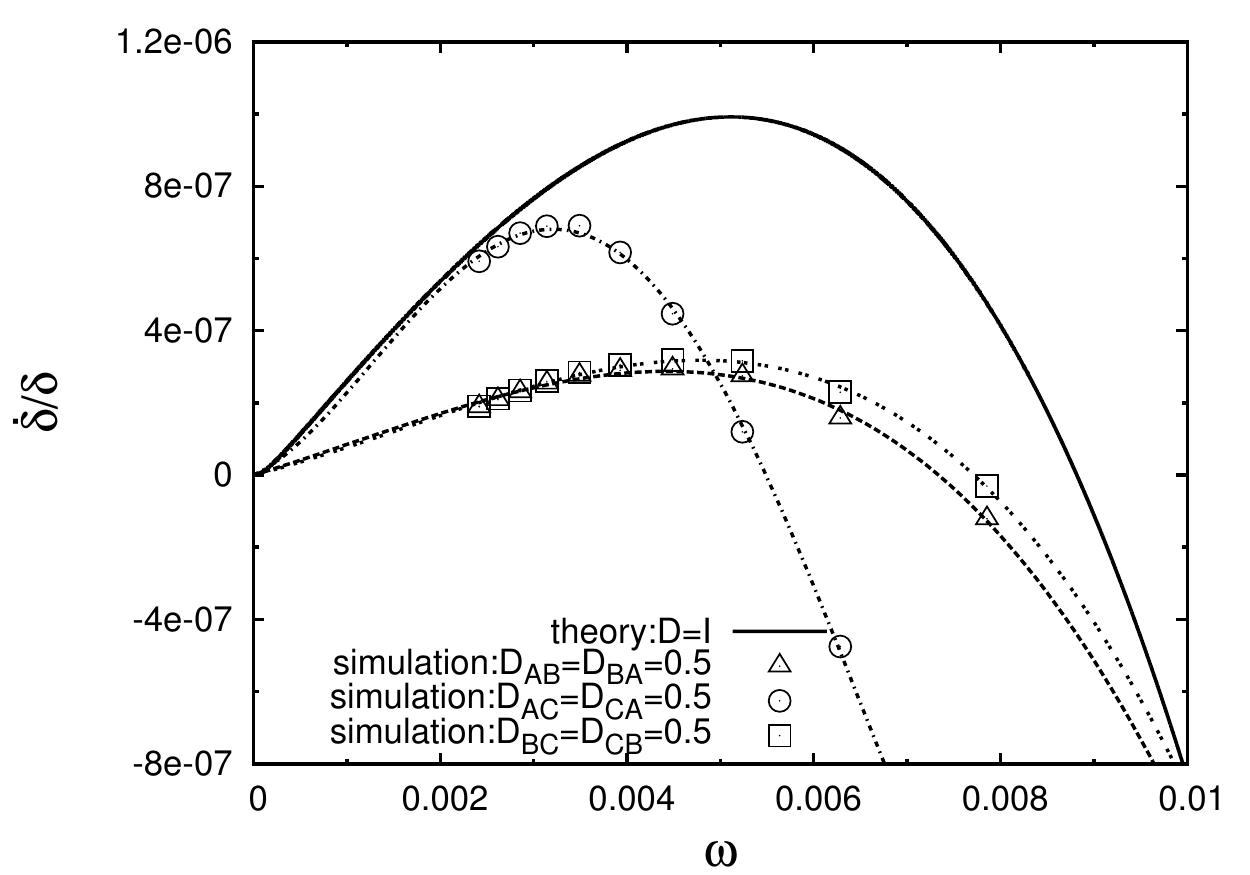}
\caption{$\dot{\delta}/\delta$ versus $\omega$ from phase-field calculations. 
The figure legend reveals the off-diagonal components which are 
non-zero with all the diagonal components set to $1$. The continuous line 
represents the analytical dispersion plot for $D=I$ and the 
discontinuous lines are polynomial functions (cubic for $D_{AB}=0.5$ and $D_{BC}=0.5$; 
quartic for $D_{AC}=0.5$) fitted to the simulation data points. The bulk liquid composition chosen here is the same as in 
Fig.~\ref{identD}.}
\label{D_off_diag}
\end{figure}
The dispersion behavior for $D_{AB}=D_{BA}=0.5$ and $D_{BC}=D_{CB}=0.5$ 
are similar, with both reporting values of $(\dot{\delta}/\delta)$ an order
of magnitude lower than that for $D=I$ with the $\omega_{max}$ (around $0.0044$ for both $D_{AB}=0.5$ and $D_{BC}=0.5$) 
taking up values similar to the case
for $D=I$ ($\omega_{max}=0.0051$). The case corresponding to $D_{AC}=D_{CA}=0.5$  
produces a behavior much different to what is observed for the other combinations. Here, for
wavenumbers smaller than $\omega_{max}$ ($=0.0032$) the values for $(\dot{\delta}/\delta)$ are very 
similar to the ones corresponding to $D=I$ but it drops 
sharply for wavenumbers larger than $\omega_{max}$ resulting in a plot whose 
$\omega_{crit}$ almost coincides with the $\omega_{max}$ for $D=I$.  

\section{Summary \& Conclusions}
Solidification of a multicomponent alloy
with unequal solute diffusivities is associated with a shift of the interfacial
compositions from the originally selected thermodynamic tie-line 
on which the alloy composition is located. These characteristics
of the transformation influence the growth behavior of morphological
perturbations and thereby influence the selection of microstructural
length scales.

In this study, we attempt to isolate and understand the effect of each of the parameters
(i.e., diffusivities, steady-state growth velocities and
tie-line compositions) on the dispersion behavior of a multicomponent alloy.
Here, we develop an analytical theory (and benchmark it against phase field simulations) 
and employ it to study the instabilities in a Hf-Re-Al-Ni 
quaternary system (A, B, C and D stand for Hf, Re, Al and Ni respectively). 

From our analytical calculations, it is observed that increase in diffusivity 
(the equilibrium tie-line and the growth velocity being kept constant) of any one of the 
components leads to a selection of larger length scales (lower $\omega_{max}$ and $\omega_{crit}$)
during the onset of instability consistent with the lower composition gradients at the interface.

The instability length scales $(2\pi/\omega_{max})$, and $(2\pi/\omega_{crit})$, are seen to reduce 
with an increase in $V$ (for a given diffusivity matrix and equilibrium compositions of the phases), 
which is explained by taking into account  
the combined influence of $V$ and the resultant planar composition 
gradients ($G_i$)'s.

Having understood the changes affected by varying diffusivity 
and the steady state velocity, we focus on 
a problem of engineering interest where 
the prediction of the dispersion behavior of an 
arbitrary alloy with known diffusivities is sought.  
A series of bulk alloy compositions along a selected 
thermodynamic tie-line choose different equilibrium compositions 
and growth velocities during steady-state growth, which can 
be correlated to the observed trends in instability length 
scales  on the basis of the concepts developed in the previous studies.

An analytical theory similar to the one for off-diagonal diffusivities is mathematically complex. 
Hence, to study the influence of the presence of cross diffusion terms in the 
diffusivity matrix on the dispersion behavior, we employ phase field simulations,
which reveal the effect of a non-zero $D_{AC}$ on the instability
length scales to be significantly different from that obtained with non-zero terms of 
$D_{BC}$ or $D_{AB}$, for the chosen system thermodynamics.

To conclude, we have been able to identify and understand the factors 
determining the length scale selection in microstructures 
arising out of Mullins-Sekerka type instabilities in multicomponent alloys. 
We have shown that our analytical expressions 
(for independent diffusion of solutes)
and phase field simulations (for both independent and coupled diffusion of solutes) 
are equally capable of predicting the instability dynamics
and length scales during solidification. This has important implications 
in casting of engineering alloys as it 
allows an efficient control on the microstructural length scales.  


\section{Acknowledgments}
We are grateful to Prof. Dipankar Banerjee and K. S. Vinay, of the Department of Materials Engineering, Indian Institute of Science, 
for bringing this particular problem to our notice and helping us with the retrieval of the relevant 
thermodynamic information from Thermo-Calc. We also 
thank Sebastian Schulz, of the Institute of Materials and Processes, Karlsruhe University of Applied Sciences, who is working on 
a similar problem for ternary systems, for critical and insightful discussions.

\section{Appendix}

\subsection{Equilibrium across a curved interface}
For determining $b_i$ Eq.~\ref{2d_equi} can be restated as,       
\begin{align} \label{loc_equi_b}
  \dfrac{1}{V_m}\sum_{i=1}^{K-1}\sum_{j=1}^{K-1} \frac{\partial \mu_i}{\partial c_j} (c_{j,\Phi}-c_{j,eq}^l)(c_{i,eq}^s-c_{i,eq}^l) &= \nonumber  \\
  \dfrac{1}{V_m}\sum_{i=1}^{K-1}\sum_{j=1}^{K-1} \frac{\partial \mu_i}{\partial c_j} (b_j \delta \sin \omega x)(c_{i,eq}^s-c_{i,eq}^l)
  &=\sigma \kappa  \\ \nonumber 
  &=\sigma \delta \omega^2 \sin \omega x,
\end{align}
where we have used Eq.~\ref{pert_interf} to obtain the second equality.

\subsection{Linearized phase diagram}
The coexistence surfaces are defined by the inter-relationships between $\mu_{i,eq}$, given by, 
\begin{align}
 \Delta\Psi^{ls} =\dfrac{1}{V_m}\left\lbrace c_{i,eq}^{s,*} - c_{i,eq}^{l,*}\right\rbrace \left\lbrace\mu_{i,eq}-\mu_{i,eq}^{*}\right\rbrace= 0.
 \label{2d_equi_pd}
\end{align}
This leads to phase compositions along the equilibrium coexistence surfaces, computed as, 
\begin{align}
  \left\lbrace c_{i,eq}^{l,s}\right\rbrace &= \left\lbrace c_{i,eq}^{l,s,*}\right\rbrace 
  +  \left[\dfrac{\partial c_i^{l,s}}{\partial \mu_j}\right]\left\lbrace \mu_{j,eq} - \mu_{j,eq}^{*}\right\rbrace.
  \label{c_of_mu_pd}
\end{align}

\subsection{Equilibrium compositions and thermodynamic parameters}
The equilibrium compositions chosen for the liquid: $c_{A}^l=0.0032$,  $c_{B}^l=0.0092$, and $c_{C}^l=0.1969$, and that for the solid 
($\gamma$):
$c_{A}^s=0.000556$, $c_{B}^s=0.023831$, $c_{C}^s=0.177447$. The data corresponds to a temperature of $1689$K.     

The $\partial c/\partial \mu$ matrices were derived by assuming a free energy density of the form:
\begin{equation}
 f=\sum_i c_i \ln c_i,
 \label{dilute_f}
 \end{equation}
which corresponds to a dilute limit approximation. Assuming, $D$ to be the solvent we compute
$\partial \mu/\partial c=\partial^2 f/\partial c ^2 $  as:

\begin{equation}
\dfrac{\partial \mu}{\partial c} = 
\left[ \begin{array}{ccc}
\frac{1}{c_A}+\frac{1}{c_D} & \frac{1}{c_D} & \frac{1}{c_D} \\
\frac{1}{c_D} & \frac{1}{c_B}+\frac{1}{c_D} & \frac{1}{c_D} \\
\frac{1}{c_D} & \frac{1}{c_D} & \frac{1}{c_C}+\frac{1}{c_D}
\end{array} \right].
\label{gen_for_dc_dmu}
\end{equation}

On substituting the equilibrium values of the components for a particular phase in Eq.~\ref{gen_for_dc_dmu}, and inverting the resultant matrix, 
we get:
\begin{equation}
{\dfrac{\partial c}{\partial \mu}}^s = 
\left[ \begin{array}{ccc}
0.000555 & -0.000013 & -0.000099  \\
-0.000013 & 0.023264 & -0.004229  \\
-0.000099 & -0.004229 & 0.145959 
\end{array} \right]
\end{equation}

\begin{equation}
{\dfrac{\partial c}{\partial \mu}}^l = 
\left[ \begin{array}{ccc}
0.00319 & -0.000029 & -0.00063  \\
-0.000029 & 0.009115 & -0.001811  \\
-0.00063 & -0.001811 & 0.15813
\end{array} \right]
\end{equation}

\subsection{Phase-field simulation parameters}
The values of the parameters controlling the interfacial energy and width 
in our phase-field simulations, are, $\sigma=0.1$ and $\epsilon=16$,
respectively. A square grid with spacing $dx=dy=4$, is used to discretize the 
dependent variable fields, and the time-stepping size is $dt=0.25$. 
The simulation box was $2000$ grid-points long, with its width being set by the 
perturbation wavelength considered.

\subsection{Non-dimensionalization}
In this study, all calculations are performed in a non-dimensionalized form. The non-dimensional
numbers can be converted back to their dimensional forms for any system, using the 
following definitions of length, time and energy scales, determined by the dimensional values
of the parameters for that particular system,
\begin{align}
 f^* &= \dfrac{1}{V_m}\left[\dfrac{\partial \mu_i}{\partial c_j}\right]_{max}, \\
 l^* &= \dfrac{\sigma}{f^*}, \\
 t^* &= \dfrac{{l^*}^2}{\left[D_{ij}\right]_{max}}.
 \label{non-dimensionalization}
\end{align}
In typical alloy solidification studies, $V_{m}=10e-06$~$m^3/mol$ and the maximum
value of diffusivity ($\left[D_{ij}\right]_{max}$) is around $1e-09$~$m^2/s$.

\subsection{Tie-line and steady-state velocity selection during growth in a multicomponent alloy}
The expressions which lead to the calculation of bulk liquid compositions ($c_i^\infty$, $i$ representing any of the $K-1$ 
independent solute elements in a $K$ component alloy), 
each corresponding to a different scaling constant $\eta_s$ ($=x_f/\sqrt{t}$, $x_f$ being the position of the solidification front at time $t$)
during steady-state solidification, 
for a given tie-line ($c_{i,eq}^l$) and independent solute diffusivities in the liquid ($D_{ii}$), 
are given as,

\begin{align}
 \left\lbrace c_i^\infty \right\rbrace &= 
 \left\lbrace c_{i,eq}^l -\dfrac{\sqrt \pi}{2}\dfrac{\eta_s \Delta c_i \erfc \left(\dfrac{\eta_s}{2\sqrt{D_{ii}}}\right)}{\sqrt{D_{ii}} \exp\left(\dfrac{-\eta_s^{2}}{4D_{ii}}\right)}\right \rbrace, \nonumber \\
 \label{composition_profiles}
\end{align}
where $\Delta c_i = c_{i,eq}^l - c_{i,eq}^s$, and $\left\lbrace \cdot \right\rbrace$ 
denotes a vector of length equal to the number of solutes in the alloy system.

We can also compute the equilibrium tie-line compositions ($c_{i,eq}^l$) and $\eta_s$ from a knowledge of
the bulk alloy compositions ($c_i^\infty$) and the independent solute diffusivities in the liquid ($D_{ii}$),
by solving a coupled set of non-linear equations given by,
\begin{align}
 \left\lbrace \dfrac{-\sqrt{\pi}\eta_s \Delta c_i}{2\sqrt{D_{ii}}} \right\rbrace &= \left\lbrace \dfrac{\left(c_i^\infty - c_{i,eq}^l\right)}{\erfc \left(\dfrac{\eta_s}{2\sqrt{D_{ii}}}\right)} \exp \left(\dfrac{-\eta_s^{2}}{4D_{ii}}\right) \right\rbrace,
 \label{eq_comp}
\end{align}
where, we employ,
\begin{align}
 \left\lbrace c_{i,eq}^{s,l} \right\rbrace =  \left\lbrace c_{i,eq}^{s,l,*} \right\rbrace + \left[ \dfrac{\partial c_i^{s,l}}{\partial \mu_j} \right] \left\lbrace \mu_{j,eq}-\mu_{j,eq}^{*} \right\rbrace, 
 \label{c_of_mu_mod}
\end{align}
and, 
\begin{align}
 \dfrac{1}{V_m}\left\lbrace c_{i,eq}^{s,*} - c_{i,eq}^{l,*}\right\rbrace \left\lbrace\mu_{i,eq}-\mu_{i,eq}^{*}\right\rbrace=0.
 \label{1d_equi}
\end{align}
to express $c_{i,eq}^{s,l}$ as functions of $\mu_{i,eq}$ (both Eqs.~\ref{c_of_mu_mod} and~\ref{1d_equi} are
expressed in the matrix-vector notation with $\left[ \cdot \right]$ denoting a matrix). In Eq.~\ref{c_of_mu_mod}, $c_{i,eq}^{s,l,*}$
represents the compositions about which the 
phase diagram is linearized, with $\mu_{j,eq}^{*}$ representing the corresponding chemical potentials. Eq.~\ref{1d_equi} is obtained
from Eq.~\ref{2d_equi} under conditions of steady-state solidification by a planar front. Thus, for $K-1$ independent solutes, Eq.~\ref{eq_comp}, 
represents same number of independent equations, with any of the $K-2$ (out of $K-1$) $\mu_i$ and $\eta_s$ as the independent 
variables to be solved for. Once, they are known,
the new equilibrium compositions $c_{i,eq}^{s,l}$ can be retrieved from Eq.~\ref{c_of_mu_mod} and Eq.~\ref{1d_equi}.

\bibliography{micro_ref}

\end{document}